\documentclass[11pt,a4paper,oneside]{article}
\usepackage{lineno,hyperref}

\usepackage{graphicx}              % to include figures
\usepackage{epsfig}
\usepackage{amsmath}               % great math stuff
\usepackage{amssymb}
\usepackage{epstopdf}
\usepackage{verbatim}
\usepackage{mathptmx}
\usepackage{mathalpha}
\usepackage{mathtools}
\usepackage[scale=0.80]{geometry}
\usepackage{graphicx, times}
\usepackage{amsfonts}              % for blackboard bold, etc
\usepackage{amsthm}                % better theorem environments\right 
\usepackage{multicol}
\usepackage{algorithm}
\usepackage{algorithmic}
\usepackage{varwidth}
\usepackage{parskip}
\usepackage{hyperref}
\usepackage{rotating}
\usepackage[numbers,sort&compress]{natbib}
\usepackage{multirow}
\usepackage{pdflscape}
\usepackage[numbers]{natbib}
\usepackage{caption}
\usepackage{xcolor}
\usepackage{color}
\usepackage{comment}
\usepackage{subcaption}
\newcommand*{\rom}[1]{\expandafter\@\romannumeral #1}
\newcommand{\bea}{\begin{eqnarray}}
	\newcommand{\eea}{\end{eqnarray}}
\newcommand{\bee}{\begin{eqnarray*}}
	\newcommand{\eee}{\end{eqnarray*}}

\begin{document}
\author{Khomesh R. Patle$^{}$\footnote{khomeshpatle5@gmail.com}, G. P. Singh$^{}$\footnote{gpsingh@mth.vnit.ac.in}
\vspace{.3cm}\\
${}^{}$ Department of Mathematics,\\ Visvesvaraya National Institute of Technology, Nagpur, 440010, Maharashtra, India.
\vspace{.3cm}
\date{}}
\title{Revisiting $f(T)$ Teleparallel Gravity with a Parametrized Hubble Parameter and Observational Constraints}
\maketitle
\begin{abstract} \noindent
In this paper, the dynamical behavior of the accelerated expansion of the universe is studied within the framework of $f(T)$ gravity by considering a well-motivated functional form of $f(T)$. A specific form of the Hubble parameter is assumed, which under two different cases, leads to two distinct cosmological models expressed in terms of the redshift parameter $H(z)$, providing insights into cosmic dynamics. These models are employed to explore the expansion history of the universe and the evolution of several cosmological parameters. Using Bayesian statistical techniques based on the $\chi^{2}$-minimization method, the median values of the model parameters are determined for both the cosmic chronometer (CC) and the joint (CC + Pantheon) datasets. The evolution of the deceleration parameter, energy density, pressure and the equation of state parameter for dark energy is analyzed. Additionally, the validity of the energy conditions and the nature of the statefinder diagnostic are examined. The present age of the universe is also estimated for the proposed models.
\end{abstract}
{\bf Keywords:} Flat FLRW Metric, $ f(T) $ Gravity, Observations, Energy Conditions, Age of the Universe. 

\section{Introduction}\label{sec:1}
Numerous independent cosmological observations have convincingly demonstrated that the universe is presently experiencing an accelerated expansion~\cite{1998AJ....116.1009R,1999ApJ...517..565P,2020A&A...641A...6P}.
This remarkable discovery has led to the introduction of dark energy as a dominant component responsible for driving this late-time acceleration. However, despite its significant contribution to the total energy density of the universe, the true nature of dark energy remains one of the most profound mysteries in modern cosmology. Current observational data suggest that dark energy together with dark matter, accounts for nearly $95$ – $96 \%$ of the total cosmic energy budget~\cite{weinberg1989cosmological}. The cosmological constant ($\Lambda$) is widely regarded as one of the simplest and most successful candidates for dark energy. While the $\Lambda$CDM model provides an excellent fit to observational data, it is plagued by two serious theoretical difficulties, including the fine-tuning problem and the coincidence problem~\cite{di2021realm,carroll2001cosmological,padmanabhan2003cosmological,copeland2006dynamics}. These unresolved issues have motivated the development of alternative cosmological models. Researchers have explored a variety of approaches, including the introduction of exotic forms of matter and energy as well as modifications to Einstein’s theory of General Relativity (GR). In particular, modified gravity theories have attracted considerable attention as promising frameworks capable of explaining cosmic acceleration without invoking an explicit dark energy component. Over the years, numerous modified gravity models have been proposed and extensively investigated in the literature~\cite{buchdahl1970non,harko2011f,nojiri2011unified,jimenez2018coincident,nojiri2017modified,bamba2010finite,elizalde2010lambdacdm,harko2010f,capozziello2019extended,capozziello2023role,kotambkar2017anisotropic,lalke2023late,hulke2020variable,garg2024cosmological,singh2025observational,patle2025accelerated,singh2024conservative,chaudhary2025extracting,shukla2025multi,goswami2024flrw,patle2026dynamical}.
\vspace{0.2cm}\\
A particularly compelling and widely investigated modification of gravity is $f(T)$ teleparallel gravity~\cite{Bengochea,cai2016f}. This theory extends the teleparallel equivalent of GR (TEGR) by generalizing the torsion scalar $T$ in the gravitational action to an arbitrary function $f(T)$. In this respect, it is conceptually analogous to $f(R)$ gravity, where the Einstein-Hilbert Lagrangian is replaced by a function of the Ricci scalar ($R$). The fundamental distinction, however lies in the geometric framework. General Relativity is formulated using the torsion-free Levi-Civita connection, where gravity is interpreted as a manifestation of spacetime curvature. In contrast, teleparallel gravity employs the curvature-free Weitzenböck connection and attributes gravitational effects to torsion rather than curvature. This alternative geometric description provides a different yet dynamically equivalent formulation at the level of TEGR, while its generalization to $f(T)$ leads to modified field equations with rich cosmological consequences. Owing to its mathematical structure and its potential to explain late-time cosmic acceleration without explicitly introducing a dark energy component, $f(T)$ gravity has attracted substantial attention. A variety of studies have explored cosmological solutions describing the dynamical evolution of the universe~\cite{paliathanasis2016cosmological} as well as its thermodynamic properties within this framework~\cite{salako2013lambdacdm}. Cosmographic methods have been applied to reconstruct the expansion history in a model-independent way~\cite{capozziello2011cosmography} and classical energy conditions have been examined to test the physical consistency of different $f(T)$ models~\cite{liu2012energy}. Moreover, alternative early-universe scenarios, including matter-bounce cosmologies have been investigated in this context~\cite{cai2011matter}. Observational analyses have further strengthened the relevance of $f(T)$ gravity. Several datasets have been used to constrain model parameters and assess their compatibility with cosmological observations~\cite{Cai}. Notably, Zhadyranova et al.~\cite{zhadyranova2024exploring} carried out a comprehensive study of late-time cosmic acceleration using a linear $f(T)$ model constrained by observational data. A detailed and systematic review of the theoretical foundations and cosmological applications of $f(T)$ gravity is presented in Ref.~\cite{cai2016f}. Subsequent influential studies include the work of Bamba et al.~\cite{bamba2011equation}, who examined the evolution of the dark energy equation of state parameter ($\omega_{de}$) in exponential, logarithmic and hybrid $f(T)$ models. Paliathanasis et al.~\cite{paliathanasis2014new} performed a complete Noether symmetry analysis, providing insights into conserved quantities and integrability properties within the theory. Furthermore, Capozziello et al.~\cite{capozziello2017model} proposed a model-independent numerical formalism for solving the modified Friedmann equations in teleparallel cosmology. Collectively, these investigations along with numerous other studies reported in Refs.~\cite{maurya2024study,shekh2025cosmographical,duchaniya2024attractor,maurya2023anisotropic,maurya2022accelerating,bamba2016bounce,bhar2024anisotropic,kavya2024can,nunes2016new,chaudhary2024constraints,duchaniya2022dynamical,mandal2020temporal,chakraborty2023classical,maurya2024role,dixit2021probe,das2023study,ren2022gaussian}, demonstrate the broad theoretical scope and observational relevance of $f(T)$ gravity as a promising framework for addressing key challenges in modern cosmology.
\vspace{0.2cm}\\
Based on the success of teleparallel gravity in addressing current cosmological issues, we are motivated to investigate a well-motivated functional form of $f(T)$ to study late-time cosmic phenomena within the framework of $f(T)$ gravity. In the present work, we consider a specific form of the Hubble parameter $H(t)$, which under two different cases, suggests two distinct cosmological models. The model parameters are constrained using the cosmic chronometer (CC) dataset and the joint (CC + Pantheon) dataset. This research work focuses on a detailed investigation of the late-time accelerated expansion of the universe in the $f(T)$ gravity model through the analysis of various cosmological parameters. 
\vspace{0.2cm}\\
The structure of this paper is organized as follows: Section \ref{sec:2} presents the basic formalism of $f(T)$ gravity and the field equations for the FLRW metric, establishing the foundation for our cosmological analysis. In Section \ref{sec:3}, we explore the parametric representations of the Hubble parameter $H(t)$ by considering two distinct cosmological models, facilitating the analysis of cosmic evolution.
Section (\ref{sec:4}), presents the constraints on model parameters using Bayesian statistical analysis, incorporating observational data from cosmic chronometers (CC) and the joint (CC+Pantheon) datasets. Section \ref{sec:5} focuses on the physical behavior and cosmic dynamics of these two models, where we examine the evolution of physical quantities (including energy density, pressure, EoS parameter), energy conditions and statefinder diagnostics. Additionally, the age of the universe is estimated for both scenarios. Finally, Section (\ref{sec:6}) summarizes the primary findings and offers concluding remarks. 
%%%%%%%%%%%%%%%%%%%%%%%%%%%%%%%%%%%%%%%%%%%%%%%%%%%%%%%%%%%%%%%%%%%%%%%%%%%%
\section{Field equations in $f(T)$ theory}\label{sec:2}
The $f(T)$ theory of gravity is a modified gravitational framework based on the torsion scalar $T$, where the geometric action is represented by an algebraic function associated with the torsion. In analogy with TG, the theory employs orthonormal tetrad fields defined in the tangent space at each point of the spacetime manifold to describe the geometry. In general, the spacetime line element can be written as
\begin{equation}{\label{1}}
ds^{2}= g_{\mu\nu}dx^{\mu}dx^{\nu}= \eta_{ij}\theta^{i}\theta^{j},
\end{equation}
with the components
\begin{equation}{\label{2}}
dx^{\mu} = e^{\mu}_{i}\theta^{i},~~~~~  \theta^{i}= e^{i}_{\mu} dx^{\mu},
\end{equation}
where $\eta_{ij}$= diag$(-1,1,1,1)$ is the metric associated with flat spacetime, while $\left\{e^{i}_{\mu}\right\}$ denote the tetrad components. These tetrads satisfy the conditions
\begin{equation}{\label{3}}
e^{~~\mu}_{i} e^{i}_{~~\nu}= \delta^{\mu}_{\nu},~~~~~  e^{~~i}_{\mu} e^{\mu}_{~~j}= \delta^{i}_{j}.
\end{equation}
The fundamental connection utilized in $f(T)$ gravity is the Weitzenböck connection~\cite{aldrovandi2012teleparallel}, which is defined as follows:
\begin{equation}{\label{4}}
	\Gamma^{\alpha}_{\mu \nu}= e_{i}^{~\alpha} \partial_{\mu} e^{i}_{~\nu}= -e^{i}_{~\mu}\partial_{\nu} e^{~\alpha}_{i}.
\end{equation}
With this connection, the components of the corresponding torsion tensor are given by~\cite{linder2010einstein}
\begin{equation}{\label{5}}
T^{\alpha}_{~\mu \nu} = -\left(\Gamma^{\alpha}_{\nu \mu}-\Gamma^{\alpha}_{\mu \nu}\right)= - e^{~\alpha}_{i} \left(\partial_{\mu} e^{i}_{~\nu} - \partial_{\nu} e^{i}_{~\mu}\right).
\end{equation}
This tensor contributes to the definition of the contorsion tensor:
\begin{equation}{\label{6}}
K^{\mu \nu}_{~\alpha} = -\frac{1}{2} \left(T^{\mu \nu}_{~\alpha} - T^{\nu \mu}_{\alpha} - T^{~\mu \nu}_{\alpha}\right),
\end{equation}
which, together with the torsion tensor, results in the tensor
\begin{equation}{\label{7}}
S^{~\mu \nu}_{\alpha} = \frac{1}{2} \left(K^{\mu \nu}_{\alpha} + \delta^{\mu}_{\alpha} T^{\lambda \nu}_{~\lambda} - \delta^{\nu}_{\alpha}T^{\lambda \mu}_{~\lambda}\right).
\end{equation}
The torsion scalar $T$ is defined as a scalar quantity obtained from the torsion tensor and $S^{~\mu \nu}_{\alpha}$, is expressed as~\cite{cai2016f,maluf2013teleparallel}
\begin{equation}{\label{8}}
T= S^{~\mu \nu}_{\alpha} T^{\alpha}_{~\mu \nu} = \frac{1}{2}T^{\alpha \mu \nu } T_{\alpha \mu \nu} + \frac{1}{2}T^{\alpha \mu \nu } T_{\nu \mu \alpha} - T^{~\alpha}_{\alpha \mu } T^{\nu \mu}_{~\nu}.
\end{equation} 
Further, the action associated with this gravitational theory is given as~\cite{Bengochea,koussour2024exploring}
\begin{equation}{\label{9}}  
	S=  \frac{1}{2\kappa^{2}}\int d^{4}xe \left[T+f(T)\right] + \int d^{4}xe L_{m},
\end{equation}
where $e$ represents the determinant of the tetrad, defined as $e= det (e^{i}_{~\mu}) = \sqrt{-g}$. Performing the variation of the action (\ref{9}) with respect to the tetrad fields yields the field equations of $f(T)$ gravity:
\begin{equation}{\label{10}}
S^{~\nu \rho}_{\mu} \partial_{\rho} T f_{TT} + [e^{-1} e^{i}_{\mu} \partial_{\rho} (ee^{~\mu}_{i} S^{~\nu \lambda}_{\alpha}) + T^{\alpha}_{~\lambda \mu} S^{~\nu \lambda}_{\alpha}] f_{T} + \frac{1}{4}\delta^{\nu}_{\mu} f = \frac{\kappa^{2}}{2} \mathit{T}_{\mu}^{\nu},
\end{equation}
where $f_{T}= \frac{\partial f}{\partial T}$, $f_{TT}= \frac{\partial^{2} f}{\partial T^{2}}$ and $\mathit{T}_{\mu}^{\nu}$ is the energy-momentum tensor defined as
\begin{equation}{\label{11}}
	\mathit{T}_{\mu}^{\nu} = (\rho + \mathit{p}) u_{\mu} u^{\nu} + p \delta^{\nu}_{\mu},
\end{equation}
where $p$ and $\rho$ represent pressure and energy density, respectively, of the ordinary matter comprising of the universe. The corresponding four-velocity field of this ordinary matter, $u^{\mu}$, satisfies the condition $u^{\mu} u_{\nu}=-1$.
\vspace{0.1cm}\\
In this work, we assume the flat FLRW metric, which is widely used to facilitate
the application of the aforementioned theory in a cosmological framework. This assumption allows for the derivation of the modified Friedmann equations. The flat FLRW spacetime metric is expressed as~\cite{linder2010einstein}
\begin{equation}{\label{12}}
	ds^{2}=-dt^{2}+a^{2}(t) \delta_{ij} dx^{i} dx^{j},
\end{equation}
where $a(t)$ is the scale factor. Accordingly, the torsion scalar for the line element (\ref{12}) is obtained as $T=-6H^{2}$.
\vspace{.1cm}\\
The Friedmann equations associated with the metric (\ref{12}) are written as~\cite{cai2016f}:
\begin{equation}{\label{13}}
6H^{2}+ 12H^{2}f_{T}+f = 2 \kappa^{2}\rho,
\end{equation}
\begin{equation}{\label{14}}
2\left(2\dot{H}+3H^{2}\right)+f+4 \left(\dot{H}+3H^{2}\right) f_{T}-48H^{2}\dot{H}f_{TT}=-2 \kappa^{2}p.
\end{equation}
Here, the symbol ``dot" represents differentiation with respect to cosmic time $t$ and $H$ stands for the Hubble parameter. The energy density $\rho$ and pressure $p$ correspond to the matter content. Setting $\kappa^{2}=1$, equations (\ref{13}) and (\ref{14}) can be rewritten as
\begin{equation}{\label{15}}
	3H^{2}= \rho + \rho_{de},
\end{equation}
\begin{equation}{\label{16}}
-2 \dot{H}-3H^{2}= p+p_{de}.
\end{equation}
Here, the energy density and pressure attributed to dark energy are specified as follows
\begin{equation}{\label{17}}
	\rho_{de}= -6H^{2}f_{T}-\frac{1}{2}f,
\end{equation}
\begin{equation}{\label{18}}
	p_{de}= \frac{1}{2}f + 2\left(\dot{H}+3H^{2}\right) f_{T}+2H \dot{f_{T}}.
\end{equation}
Employing equations (\ref{17}) and (\ref{18}), we obtain the expression of the EoS parameter of dark energy as
\begin{equation}{\label{19}}
\omega_{de}=\frac{p_{de}}{\rho_{de}} = -1-\frac{4 \left(H \dot{f_{T}} + \dot{H} f_{T}\right)}{f+ 12H^{2}f_{T}}.
\end{equation}
%%%%%%%%%%%%%%%%%%%%%%%%%%%%%%%%%%%%%%%%%%%%%%%%%%%%%%%%%%%%
\section{Parametric representations of the Hubble parameter $H(z)$ }\label{sec:3}
The Hubble parameter addresses the expansion rate of the universe and plays a fundamental role in cosmological models constructed within different gravitational frameworks. Nevertheless, the cosmic expansion history can also be investigated using a model-independent approach that does not rely on any specific gravitational theory~\cite{shafieloo2013model}. In such an approach, the field equations are analyzed through cosmological parameterizations, where the key unknown quantities include the Hubble parameter, pressure and energy density. The transition of the universe from an early decelerated phase to the present accelerated phase can be effectively studied by parameterizing cosmological quantities such as the Hubble parameter, the deceleration parameter and the equation of state (EoS) parameter. These parameterizations can further be tested and constrained using observational datasets. Such methods have been widely examined in the literature to characterize significant cosmological challenges, including the initial singularity problem, the issue of an always decelerating expansion, the horizon problem and the Hubble tension, among other key issues~\cite{banerjee2005acceleration,cunha2008transition,escamilla2022dynamical}. In this context, the late-time accelerated expansion of the universe can be described by introducing an appropriate functional form of the Hubble parameter $H(z)$~\cite{myrzakulov2023quintessence,yadav2024reconstructing}. Motivated by this idea, we consider a specific parameterized form of the Hubble parameter $H$~\cite{pacif2017reconstruction} as an explicit function of cosmic time $t$ in the following form
\begin{equation}{\label{20}}
	H(t)= \frac{\tau_{2} t^{n}}{(t^{m}+\tau_{1})^{\sigma}},
\end{equation}
where $\tau_{1}$, $\tau_{2}$ $\neq 0$ and $m$, $n$, $\sigma$ are real constants (model parameters). The parameters $\tau_{1}$ and $\tau_{2}$ both have the dimensions of time. Certain specific values of the parameters $m$, $n$ and $\sigma$ lead to some distinguished cosmological models. In this paper, we consider models obtained for some integral and non-integral values of $m$, $n$ and $\sigma$ in the functional form of the Hubble parameter given in Eq. (\ref{20}). Among these, two models corresponding to ($m=1$, $\sigma=1$, $n=-1$) and ($m=2$, $\sigma=1$, $n=-1$) exhibit the possibility of describing the phenomena of cosmological phase transition for negative $\tau_{1}$ \& $\tau_{2}$. These models are presented in Table (\ref{table:1}). Here $\zeta$ is an integrating constant that also plays an important role in the cosmic evolution. In the present work, we investigate two models, Model-I and Model-II, for which a detailed analysis has been carried out.
%%%%%%%%%%%%%%%%%%%%%%%%%%%%%%%%%%%%%%%%%%%%%%%%%%%%%%%%%%%%%%%%%%%%%
\begin{table}[h!]
	\centering
	\caption{The models}
	\label{table:1}
    \renewcommand{\arraystretch}{2.5}  % Default 1.0 hai, isko badhayein
    \fontsize{16pt}{7pt}\selectfont  % {font size}{line spacing}
		\begin{tabular}{|c|c|c|c|}
			\hline
			Models & $H(t)$ & $a(t)$ & $q(t)$ \\
			\hline
			Model-I & $\dfrac{\tau_2}{t(\tau_1-t)}$ & $\zeta \left(\dfrac{t}{\tau_1-t}\right)^{\frac{\tau_2}{\tau_1}}$ & $-1 + \dfrac{\tau_1}{\tau_2} - \dfrac{2}{\tau_2} t$ \\
			\hline
			Model-II & $\dfrac{\tau_2}{t(\tau_1-t^2)}$ & $\zeta \left(\dfrac{t^2}{\tau_1-t^2}\right)^{\frac{\tau_2}{2 \tau_1}}$ & $-1 + \dfrac{\tau_1}{\tau_2} - \dfrac{3}{\tau_2} t^2$ \\
			\hline
	\end{tabular} 
\end{table}
%%%%%%%%%%%%%%%%%%%%%%%%%%%%%%%%%%%%%%%%%%%%%%%%%%%%%%%%%%%%%%%%%%%%%
\vspace{.2cm}\\
One can see that, for both models, the Hubble parameter and the scale factor diverge at a finite cosmic time, indicating the occurrence of a Big Rip singularity in the near future. For Model-I, the singularity occurs at $t$ = $t_{s}$ = $\tau_{1}$ while for Model-II, it appears at $t$ = $t_{s}$ = $\sqrt{\tau_{1}}$. Furthermore, the transition from the decelerating phase to the accelerating phase of the universe takes place at $t_{tr}$ = $\frac{\tau_{1}-\tau_{2}}{2}$ for Model-I and at $t_{tr}$ = $\sqrt{\frac{\tau_{1}-\tau_{2}}{3}}$ for Model-II. These relations imply that the condition $\tau_{1} > \tau_{2}$ must be satisfied for the transition time to be physically meaningful.
\vspace{.2cm}\\
In observational cosmology, it is convenient to express all cosmological parameters in terms of the redshift ($z$). Since the cosmological parameters considered here are functions of the cosmic time ($t$), it is necessary to establish a relation between time and redshift. The corresponding $t$-$z$ relations are obtained as follows
\begin{equation}{\label{21}}
	t(z)= \tau_{1}\left[1+(\zeta (1+z))^{\frac{\tau_{1}}{\tau_{2}}} \right]^{-1}  ~~~\qquad \qquad \qquad \qquad     \text{(for Model-I)}
\end{equation}
\begin{equation}{\label{22}}
	t(z)= \sqrt{\tau_{1}} \left[1+(\zeta (1+z))^{2\frac{\tau_{1}}{\tau_{2}}} \right]^{-\frac{1}{2}}   \qquad \qquad \qquad \qquad   \text{(for Model-II)}
\end{equation}
\vspace{0.1cm}\\
The above expressions (\ref{21}) and (\ref{22}) involve three parameters, namely $\zeta$, $\tau_{1}$ and $\tau_{2}$. However, these models can be effectively described using only two independent parameters by defining the ratio $\gamma$ = ($\tau_{1}$/$\tau_{2}$), which also simplifies the subsequent analysis. Therefore, the expressions for the Hubble parameter for both models in terms of the redshift ($z$) are obtained as:
\begin{equation}{\label{23}}
	H(z)= H_{0}(1+\zeta^{\gamma})^{-2} (1+z)^{-\gamma} \left[1+(\zeta (1+z))^{\gamma} \right]^{2}     \qquad \qquad \qquad \qquad     \text{(for Model-I)}
\end{equation}
\begin{equation}{\label{24}}
	H(z)= H_{0}(1+\zeta^{2\gamma})^{-\frac{3}{2}} (1+z)^{-2\gamma} \left[1+(\zeta (1+z))^{2\gamma} \right]^{\frac{3}{2}}  ~~~\qquad \qquad \qquad      \text{(for Model-II)}
\end{equation}
\vspace{0.1cm}\\
where, the Hubble parameter at redshift $ z=0 $ is denoted by $H_{0}$, while $\gamma$ and $\zeta$ are free model parameters that are constrained using observational data. The deceleration parameter $q$ is one of the key quantities used to describe the expansion dynamics of the universe. It provides information about the nature of cosmic expansion. Different values of the deceleration parameter characterize different phases of the universe, where $q<0$ corresponds to an accelerated expansion phase and $q>0$ represents a decelerated expansion phase. When the value of the deceleration parameter becomes smaller than $-1$, the universe enters a super-accelerated expansion regime. Observational studies indicate that specific values of $q$, namely $-1$, $\frac{1}{2}$ and $1$, correspond to the de Sitter phase, matter-dominated era and radiation-dominated era of the universe, respectively. In addition, the general form of the deceleration parameter $q(z)$ is expressed as
\begin{equation}{\label{25}}
	q(z)= -\frac{\dot{H}}{H^{2}}-1 .
\end{equation} 
A relation for the first derivative of the Hubble parameter with respect to cosmic time can be expressed as a function of redshift: $\dot{H}= -(1+z) H(z)\frac{dH(z)}{dz}$. By substituting equations (\ref{23}) and (\ref{24}) into equation (\ref{25}), the deceleration parameter $q(z) $ can be expressed within the framework of the present models as,
\begin{equation}{\label{26}}
	q(z)= -1 + \gamma \left[\frac{\left(\zeta(1+z)\right)^{\gamma}-1}{1+\left(\zeta(1+z)\right)^{\gamma}}\right]      \qquad \qquad \qquad \qquad     \text{(for Model-I)}
\end{equation}
\begin{equation}{\label{27}}
	q(z)= -1 + \gamma \left[\frac{\left(\zeta(1+z)\right)^{2\gamma}-2}{1+\left(\zeta(1+z)\right)^{2\gamma}}\right]      \qquad \qquad \qquad \qquad     \text{(for Model-II)}
\end{equation}
%%%%%%%%%%%%%%%%%%%%%%%%%%%%%%%%%%%%%%%%%%%%%%%%%%%%%%%%%%%%%%%%%%%%%%%
\begin{center}
	\begin{figure}
		\includegraphics[width=16cm, height=6cm]{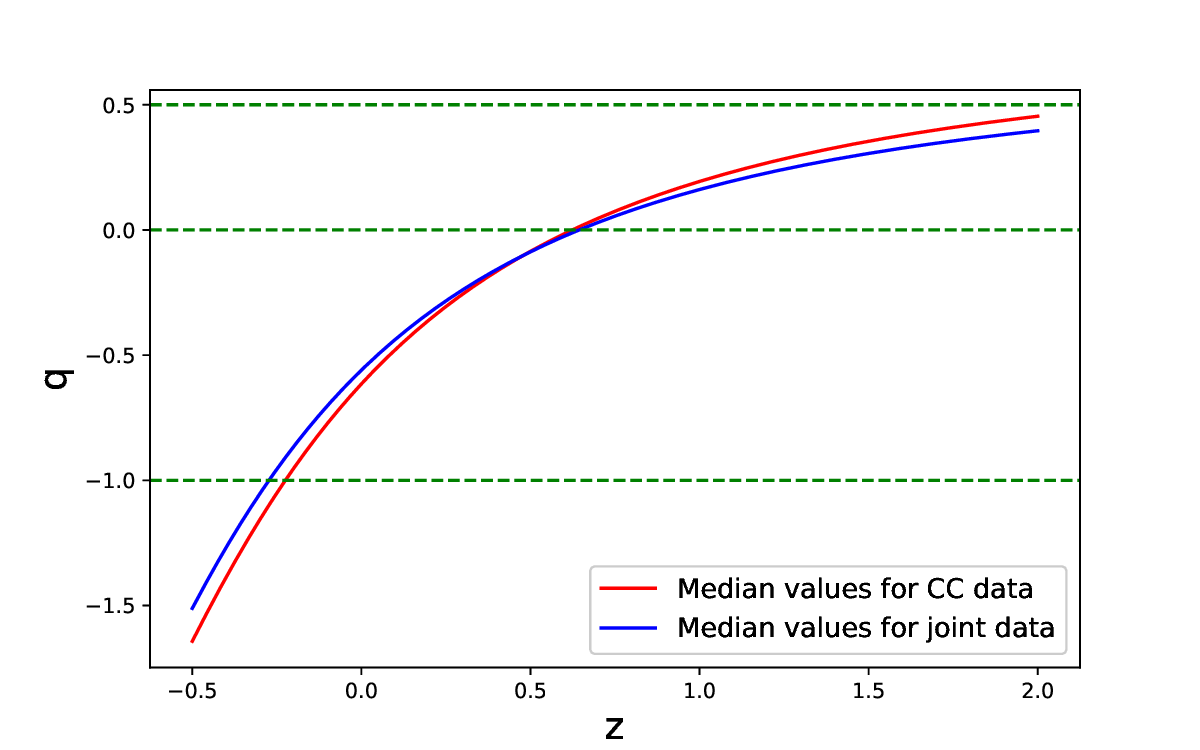}
		\caption{\textbf{For Model-I:} The deceleration parameter$(q)$ versus $z$.} 
		\label{fig:1}
	\end{figure}
\end{center}
%%%%%%%%%%%%%%%%%%%%%%%%%%%%%%%%%%%%%%%%%%%%%%%%%%%%%%%%%%%%%%%%
\begin{center}
	\begin{figure}
		\includegraphics[width=16cm, height=6cm]{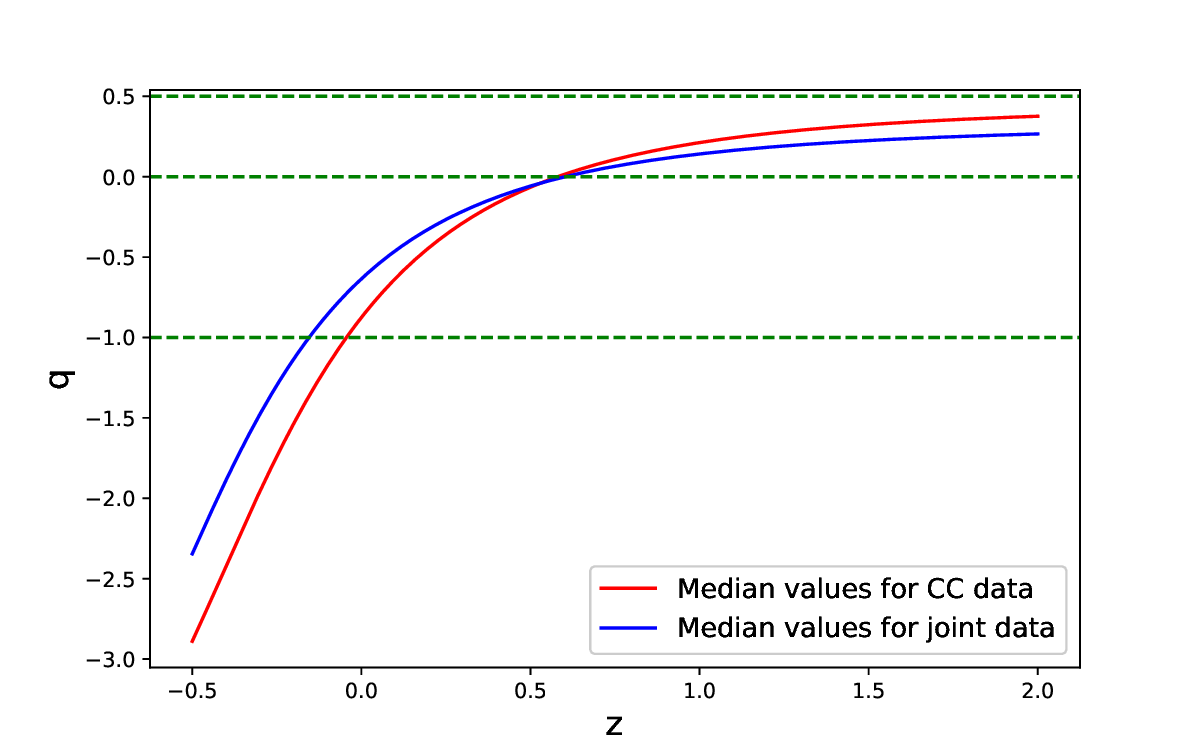}
		\caption{\textbf{For Model-II:} The deceleration parameter$(q)$ versus $z$.} 
		\label{fig:2}
	\end{figure}
\end{center}
%%%%%%%%%%%%%%%%%%%%%%%%%%%%%%%%%%%%%%%%%%%%%%%%%%%%%%%%%%%%%%%%%
The evolution of the deceleration parameter obtained from the CC dataset and the joint dataset is shown in Figures~$(\ref{fig:1})$ and $(\ref{fig:2})$. These figures clearly illustrate the transition of the universe from an earlier decelerated expansion phase to the present accelerated expansion phase. Using the median values of the model parameters, the present value of the deceleration parameter for Model-I is estimated to be $q_{0}=-0.6151$ (CC) and $q_{0}=-0.5785$ (joint) dataset. For Model-II, the corresponding present values are $q_{0}=-0.8758$ and $q_{0}=-0.6508$ for CC and joint datasets, respectively. The negative values of $q_{0}$ indicate that the universe is currently experiencing accelerated expansion at $z=0$. Furthermore, the deceleration parameter confirms the existence of the present accelerating phase of cosmic expansion. For Model-I, the transition from decelerated to accelerated expansion occurs at redshifts $z=0.6261$ and $z=0.6263$ for the CC and joint datasets, respectively. Similarly, for Model-II, the transition redshift is $z=0.6036$ (CC dataset) and $z=0.6039$ (joint dataset).
\\ \\
From the constraints obtained with both datasets, we infer that the deceleration parameter approaches $q = \frac{1}{2}$ at high redshifts, signifying that both cosmological models evolve through a matter-dominated past. Subsequently, the models predict a transition to super-exponential acceleration at late times, driven by phantom-like dark energy. Since this accelerated phase goes beyond the de Sitter limit, the underlying dark energy component is identified as quintom dark energy, as outlined in Ref.~\cite{Zhao2006}.
%%%%%%%%%%%%%%%%%%%%%%%%%%%%%%%%%%%%%%%%%%%%%%%%%%%%%%%%%%%%%%%%%%%%%%%%%%%%%%%
\section{Observational constraints on model and results}\label{sec:4}
In this section, we employ a Bayesian statistical framework to assess the consistency of the proposed cosmological model with observational datasets. To constrain the cosmological parameters $H_{0}$, $\gamma$ and $\zeta$, appearing in the parameterized Hubble parameter given in Eqs. (\ref{23}) and (\ref{24}), we utilize observational data from the cosmic chronometer (CC) sample as well as the joint (CC+Pantheon) dataset. Model fitting is performed through $\chi^{2}$ minimization, coupled with the Markov chain Monte Carlo (MCMC) sampling method, implemented using the emcee Python library~\cite{foreman2013emcee}.
\subsection{The Cosmic chronometer dataset}\label{sec:4.1}
\begin{center}
	\begin{figure}
		\includegraphics[width=16cm, height=6cm]{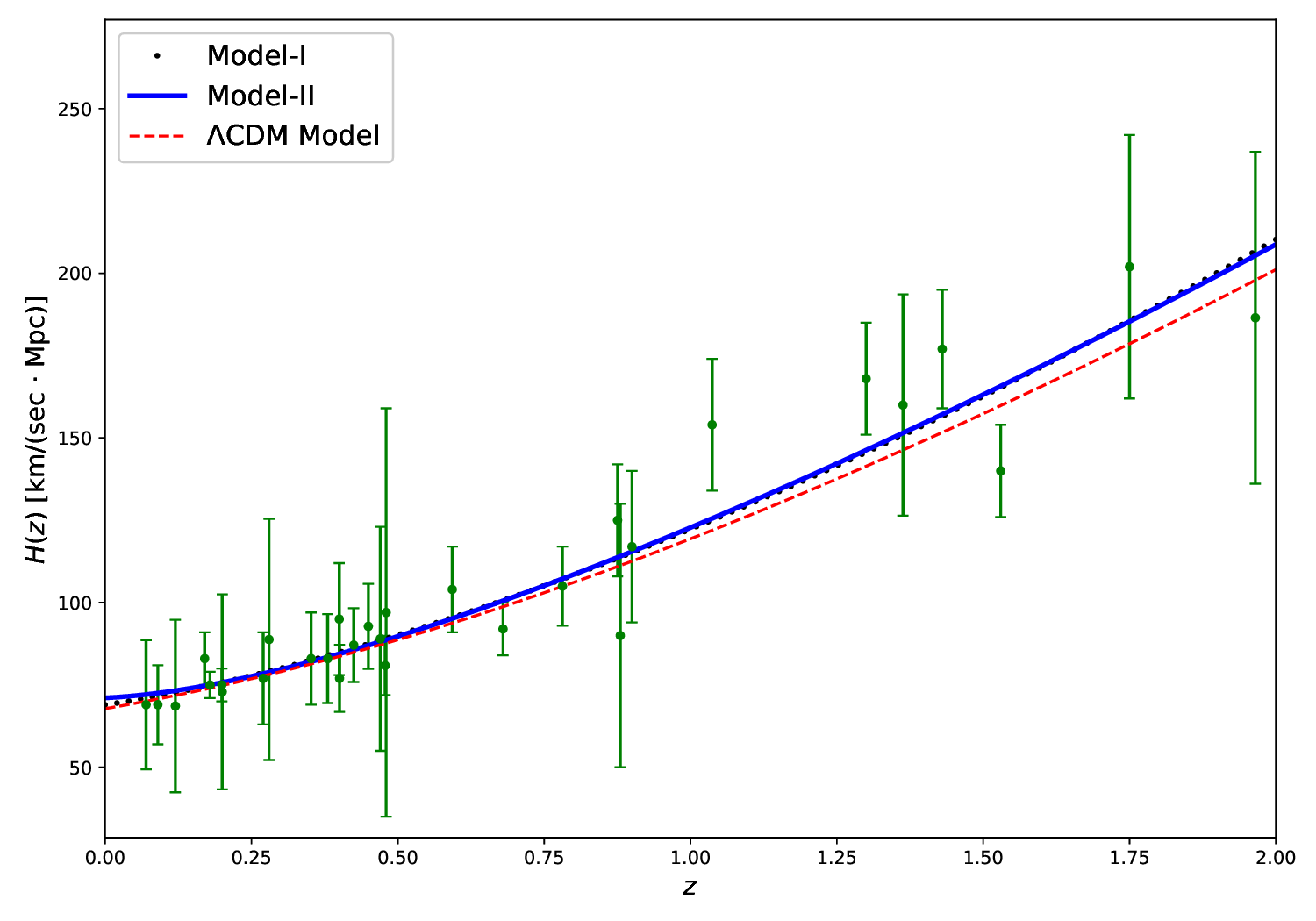}
		\caption{The best-fit $H(z)$ curve for the proposed models is compared with the $\Lambda CDM$ model.} 
		\label{fig:3}
	\end{figure}
\end{center}

In order to constrain model parameters, we proceed to scrutinize the observational viability of our proposed cosmological scenario. The analysis incorporates a dataset consisting of $31$ cosmic chronometer (CC) measurements \cite{simon2005constraints, sharov2018predictions}, acquired through the differential age (DA) technique applied to galactic systems across the redshift range $0.07 \leq z \leq 1.965$ \cite{stern2010cosmic, moresco2015raising}. The primary goal of this examination is to compute the median values characterizing the model parameters. According to the foundational methodology advanced by Jimenez and Loeb \cite{jimenez2002constraining}, the Hubble parameter $(H)$ is related to cosmic time $(t)$ and redshift $(z)$ through the fundamental relation: $H(z) = \frac{-1}{(1+z)} \frac{dz}{dt}$. To constrain the free parameters $H_{0}$, $\gamma$ and $\zeta$, we employ a conventional statistical methodology that involves minimizing the chi-squared $(\chi^{2})$ function, which is formally equivalent to the maximization of the likelihood function \cite{mandal2023cosmic,garg2025cosmological}.
\begin{equation}{\label{28}}
	\chi^{2}_{CC}(\theta)=\sum_{i=1}^{31} \frac{[H_{th}(\theta,z_{i})-H_{obs}(z_{i})]^{2} }{ \sigma^{2}_{H(z_{i})}},  
\end{equation} 
\vspace{0.1cm}\\
here $H_{th}$ stands for the theoretical prediction of the Hubble parameter, $H_{obs}$ refers to the value obtained from observations and $\sigma_{H}$ quantifies the corresponding to the standard deviation from observations.
\vspace{0.1cm}\\
Figure $(\ref{fig:3})$ illustrates the error bars for the CC data points along with the best fit Hubble parameter curve.
%%%%%%%%%%%%%%%%%%%%%%%%%%%%%%%%%%%%%%%%%%%%%%%%%%%%%%%%%%%
%%%%%%%%%%%%%%%%%%%%%%%%%%%%%%%%%%%%%%%%%%%%%%%%%%%%%%%%%%%
%%%%%%%%%%%%%%%%%%%%%%%%%%%%%%%%%%%%%%%%%%%%%%%%%%%%%%%%%%%%%%%%%%%
\subsection{The Pantheon dataset}\label{sec:4.2}
\begin{center}
	\begin{figure}
		\includegraphics[width=18.5cm, height=18.5cm]{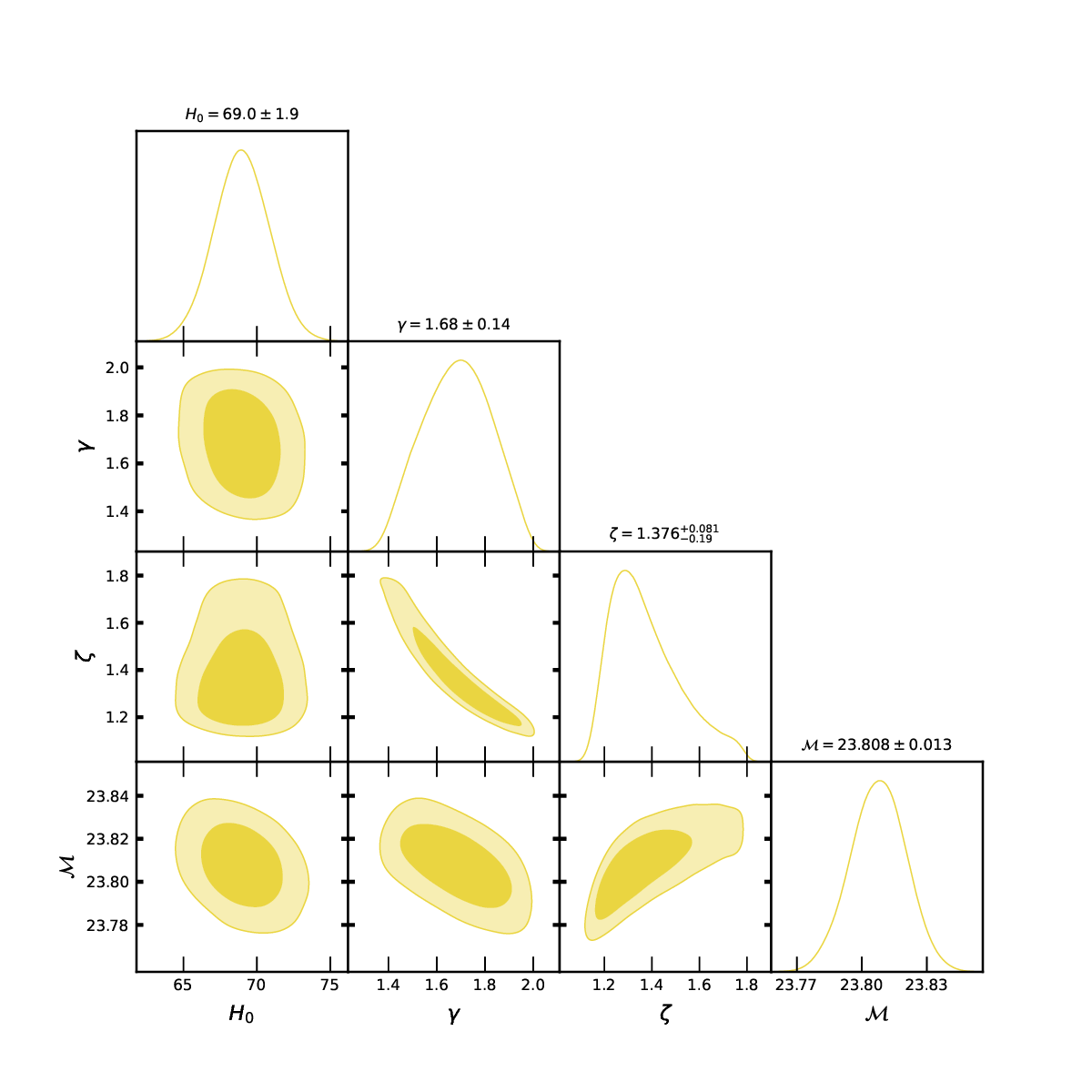}
		\caption{\textbf{For Model-I:} Marginalized $1D$ and $2D$ posterior contour map with median values of $H_{0}$, $\gamma$ and $\zeta$ using the Joint dataset.}
		\label{fig:4}
	\end{figure}
\end{center}

\begin{center}
	\begin{figure}
		\includegraphics[width=18.5cm, height=18.5cm]{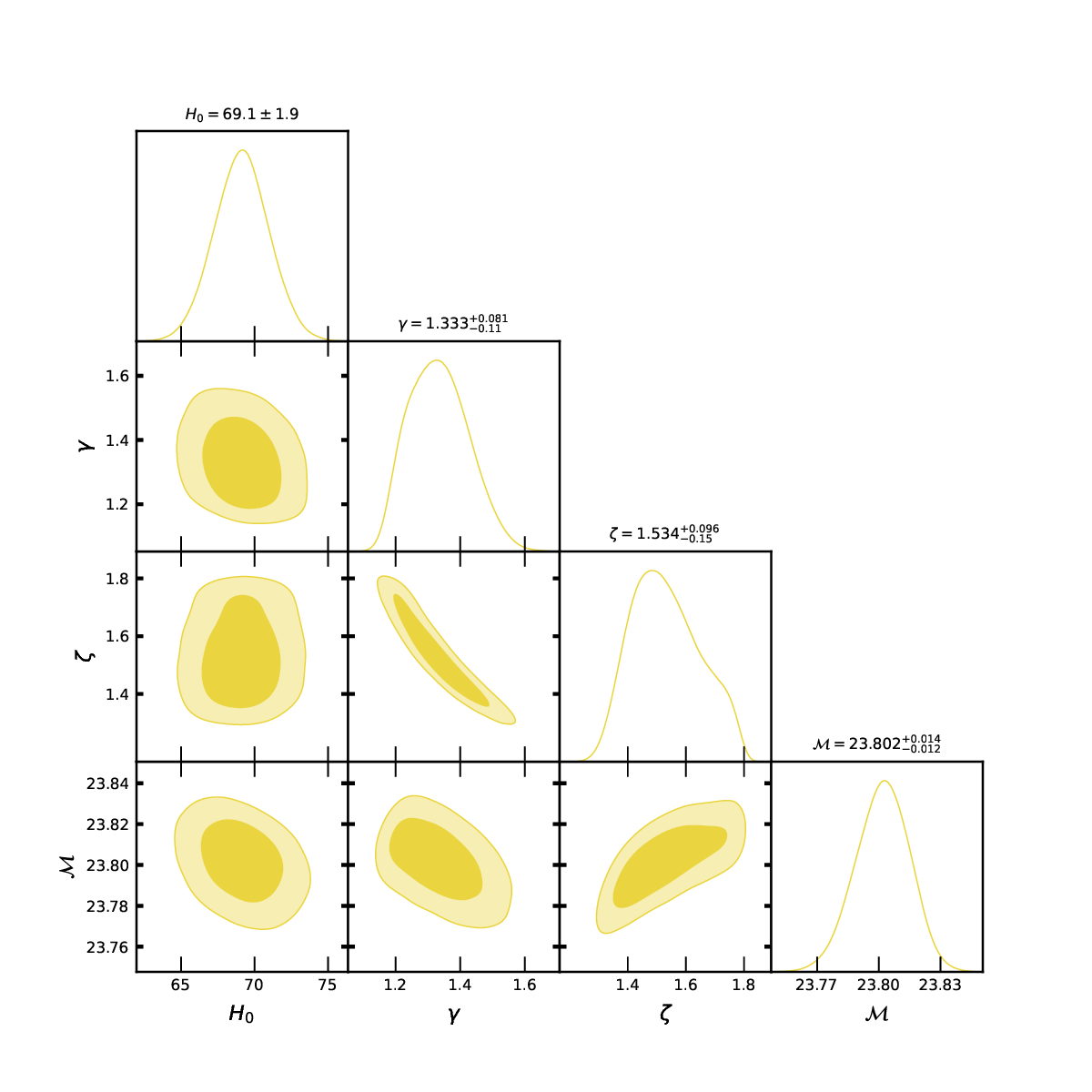}
		\caption{\textbf{For Model-II:} Marginalized $1D$ and $2D$ posterior contour map with median values of $H_{0}$, $\gamma$ and $\zeta$ using the Joint dataset.}
		\label{fig:5}
	\end{figure}
\end{center}

The present investigation employs the Pantheon compilation, which contains 1048 Type Ia supernovae (SNIa) data points spanning the redshift range $0.01 < z < 2.26$~\cite{scolnic2018complete}. This rich dataset synthesizes observations from several distinguished surveys, including the CfA1–CfA4 series~\cite{riess1999bvri, hicken2009improved}, the Pan-STARRS1 Medium Deep Survey~\cite{scolnic2018complete}, SDSS~\cite{sako2018data}, SNLS~\cite{guy2010supernova} and the Carnegie Supernova Project (CSP)~\cite{contreras2010carnegie}. When performing the MCMC analysis with the Pantheon dataset, the theoretically predicted apparent magnitude $\mu_{th}(z)$ is formulated as
\begin{equation}{\label{29}}
\mu_{th}(z)=M+5log_{10}\left[\frac{d_{L}(z)}{Mpc}\right]+25,
\end{equation}
here the symbol $M$ denotes the absolute magnitude. Meanwhile, the luminosity distance $d_{L}(z)$, which carries units of length, takes the form~\cite{odintsov2018cosmological}
\begin{equation}{\label{30}}
	d_{L}(z)=c(1+z)\int_{0}^{z}\frac{dz'}{H(z')},
\end{equation}
In this formulation, the variable $z$ denotes the redshift of Type Ia supernovae (SNIa) measured in the cosmic microwave background (CMB) rest frame and $c$ represents the speed of light. The luminosity distance $d_L(z)$ is frequently expressed in terms of its dimensionless, Hubble-free counterpart, defined as $D_{L}(z) \equiv H_{0}d_{L}(z)/c$. Consequently, equation (\ref{29}) can be recast in the following form:
\begin{equation}{\label{31}}
	\mu_{th}(z)=M+5log_{10}\left[D_{L}(z)\right]+5log_{10}\left[\frac{c/H_{0}}{Mpc}\right]+25. 
\end{equation}
There will be a degeneracy between $M$ and $H_{0}$ in the $\Lambda$CDM model framework~\cite{ellis2012relativistic,asvesta2022observational}. We express $\mathcal{M}$ as a combination of these parameters as shown below.
\begin{equation}{\label{32}}
	\mathcal{M}\equiv M+5log_{10} \left[\frac{c/H_{0}}{Mpc}\right]+25=M+42.38-5log_{10}(h), 
\end{equation}
with $H_{0}=h \times 100$ $[\text{km}/(\text{sec}.\text{Mpc})]$, we proceed with the MCMC analysis by incorporating these parameters and the corresponding $\chi^{2}$ function for the Pantheon data, as formulated in~\cite{garg2025cosmological,asvesta2022observational}:
\begin{equation}{\label{33}}
	\chi^{2}_{P}= \nabla \mu_{i}C^{-1}_{ij}\nabla \mu_{j}.
\end{equation}
Here, $\nabla \mu_{i} = \mu_{obs}(z_{i}) - \mu_{th}(z_{i})$, where $C_{ij}^{-1}$ denotes the inverse of the covariance matrix and $\mu_{th}$ is specified by equation (\ref{31}). The luminosity distance depends on the behavior of the Hubble parameter. For our analysis, we implemented the emcee package~\cite{foreman2013emcee} together with the relevant equation to perform maximum likelihood estimation (MLE) using the joint (CC+Pantheon) sample. The joint $\chi^{2}$ statistic formulated for this purpose takes the form $\chi^{2}_{CC} + \chi^{2}_{P}$. Figures $(\ref{fig:4})$ and $(\ref{fig:5})$ are exhibited for $1\sigma$ and $2\sigma$ likelihood contour maps and corresponding 1D posterior distributions derived from the MCMC sampling of the CC+Pantheon joint datasets. The resulting median values of model parameters from the MCMC analysis are provided in Table(\ref{table:2}) and Table(\ref{table:3}).
%%%%%%%%%%%%%%%%%%%%%%%%%%%%%%%%%%%%%%%%%%%%%%%%%%%%%%%%%%%%%%%%%%%%%%
\begin{table}[htbp]
	\centering
	\renewcommand{\arraystretch}{2.5}  
	\fontsize{9pt}{9pt}\selectfont  % {font size}{line spacing}
	\begin{tabular}{|c|c|c|c|c|c|c|c|c|c|}
		\hline
		Dataset & $H_{0}$[Km/(sec.Mpc)] & $\gamma$ & $\zeta$ & $\mathcal{M}$ & $q_{0}$ & $z_{t}$ & $\omega_{0}$ & $t_{0}$(Gyr) \\
		\hline
		CC & $68.956^{+0.950}_{-0.931}$ & $1.752^{+0.067}_{-0.068}$ & $1.290^{+0.098}_{-0.078}$ & - & $-0.6151$ & $0.6261$ & $-0.6313$ & $13.27$ \\
		\hline
		CC+Pantheon  & $69.0^{+1.9}_{-1.9}$  & $1.68^{+0.14}_{-0.14}$ &  $1.376^{+0.081}_{-0.19}$ &  $23.808^{+0.013}_{-0.013}$ & $-0.5785$ & $0.6263$ & $-0.5785$ & $13.38$ \\
		\hline
	\end{tabular} 
	\caption{ \textbf{For Model-I:} For both CC and joint datasets, the median values of the model parameters, along with the present values of $q_{0}$, $\omega_{0}$ and $t_{0}$.}
	\label{table:2}
\end{table}
%%%%%%%%%%%%%%%%%%%%%%%%%%%%%%%%%%%%%%%%%%%%%%%%%%%%%%%%%%%
%%%%%%%%%%%%%%%%%%%%%%%%%%%%%%%%%%%%%%%%%%%%%%%%%%%%%%%%%%%%%%%%%%%%%%
\begin{table}[htbp]
	\centering
	\renewcommand{\arraystretch}{2.5}  
	\fontsize{9pt}{9pt}\selectfont   %{font size}{line spacing}
	\begin{tabular}{|c|c|c|c|c|c|c|c|c|c|}
			\hline
			Dataset & $H_{0}$[Km/(sec.Mpc)] & $\gamma$ & $\zeta$ & $\mathcal{M}$ & $q_{0}$ & $z_{t}$ & $\omega_{0}$ & $t_{0}$(Gyr) \\
			\hline
			CC & $71.063^{+1.114}_{-1.110}$ & $1.452^{+0.039}_{-0.041}$ & $1.328^{+0.064}_{-0.052}$ & - & $-0.8758$ & $0.6036$ & $-0.8810$ & $13.63$ \\
			\hline
			CC+Pantheon  & $69.1^{+1.9}_{-1.9}$  & $1.333^{+0.081}_{-0.11}$ &  $1.534^{+0.096}_{-0.15}$ &  $23.802^{+0.014}_{-0.012}$ & $-0.6508$ & $0.6039$ & $-0.6508$ & $14.01$ \\
			\hline
	\end{tabular} 
     \caption{ \textbf{For Model-II:} For both CC and joint datasets, the median values of the model parameters, along with the present values of $q_{0}$, $\omega_{0}$ and $t_{0}$.}
	\label{table:3}
\end{table}
%%%%%%%%%%%%%%%%%%%%%%%%%%%%%%%%%%%%%%%%%%%%%%%%%%%%%%%%%%%
%%%%%%%%%%%%%%%%%%%%%%%%%%%%%%%%%%%%%%%%%%%%%%%%%%%%%%%%%%%
%%%%%%%%%%%%%%%%%%%%%%%%%%%%%%%%%%%%%%%%%%%%%%%%%%%%%%%%%%%%%%%%%%%%
%%%%%%%%%%%%%%%%%%%%%%%%%%%%%%%%%%%%%%%%%%%%%%%%%%%
\begin{center}
	\begin{figure}
		\includegraphics[width=16cm,
		height=6cm]{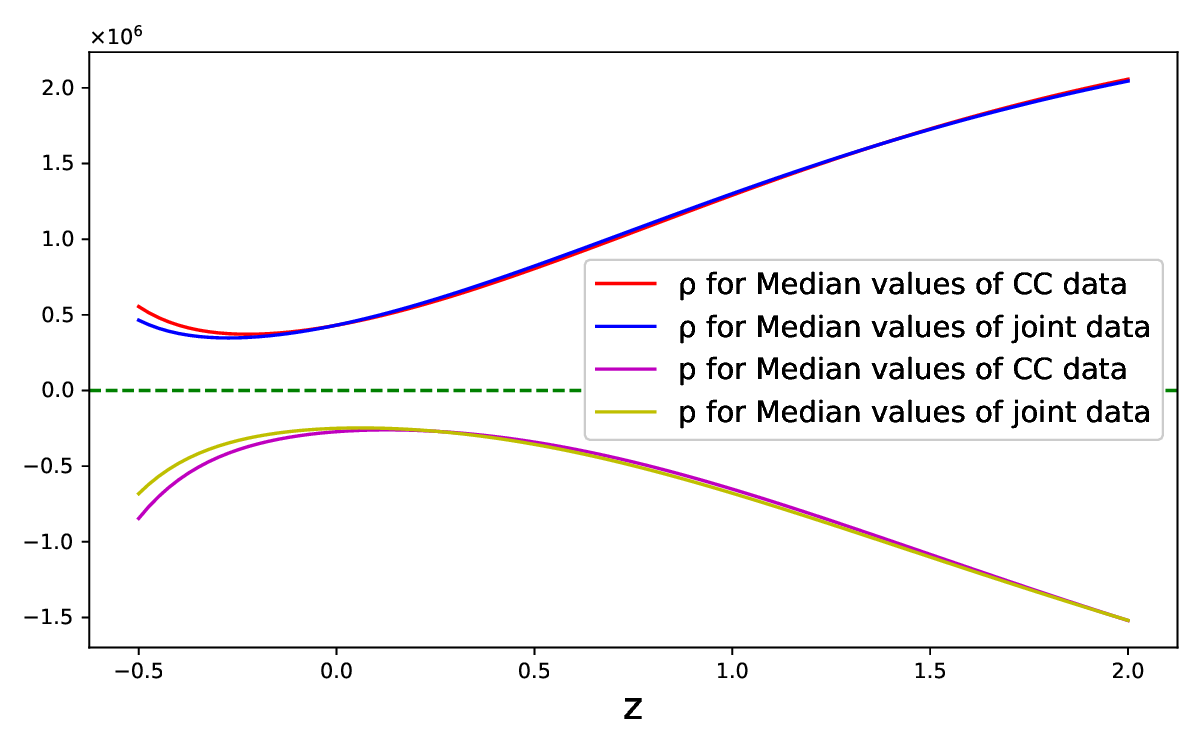}
		\caption{ {\bf{For Model-I:}} $\rho_{de}$ and $p_{de}$ versus $\mathit{z}$.}
		\label{fig:6}
	\end{figure}
\end{center}
%%%%%%%%%%%%%%%%%%%%%%%%%%%%%%%%%%%%%%%%%%%%%%%%%%%
%%%%%%%%%%%%%%%%%%%%%%%%%%%%%%%%%%%%%%%%%%%%%%%%%%%%%%%%%%%
\section{Physical Behavior and Cosmic Dynamics of the Cosmological Models}\label{sec:5}
\subsection{Analyzing the Physical Behavior of the Models}\label{subsec:5.1}
\begin{center}
	\begin{figure}
		\includegraphics[width=16cm, height=6cm]{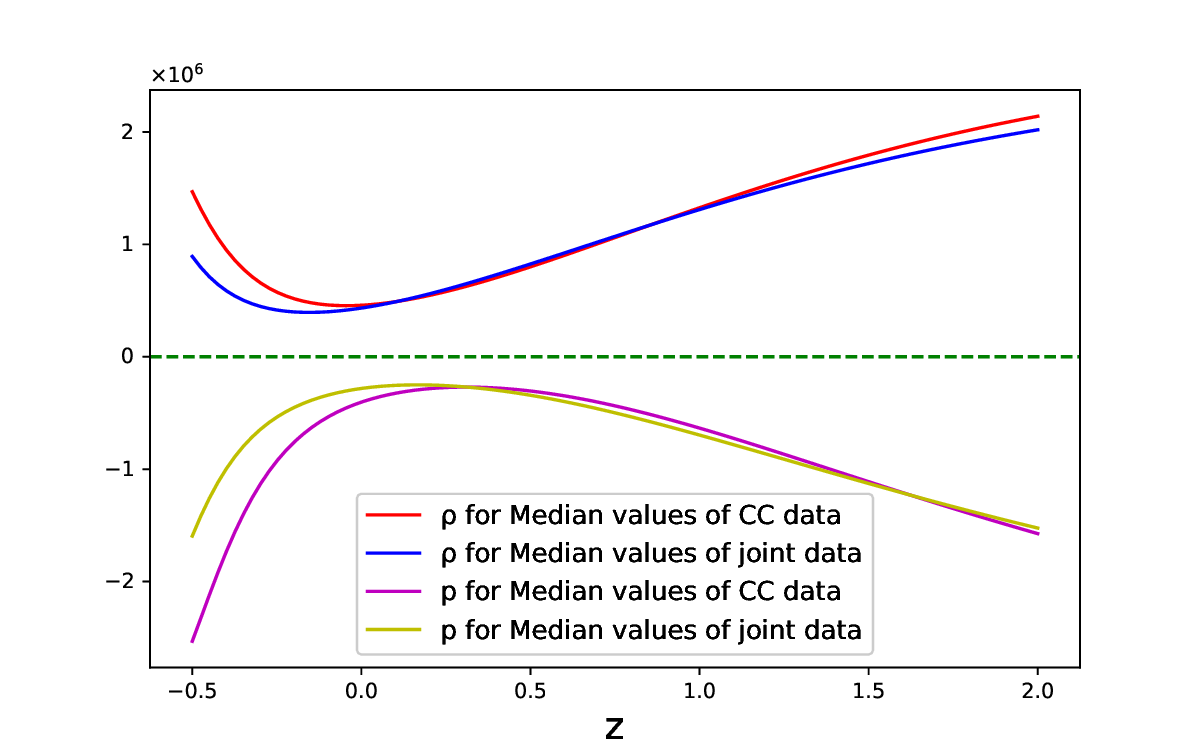}
		\caption{ {\bf{For Model-II:}} $\rho_{de}$ and $p_{de}$ versus $\mathit{z}$.}
		\label{fig:7}
	\end{figure}
\end{center}

\begin{center}
	\begin{figure}
		\includegraphics[width=16cm, height=6cm]{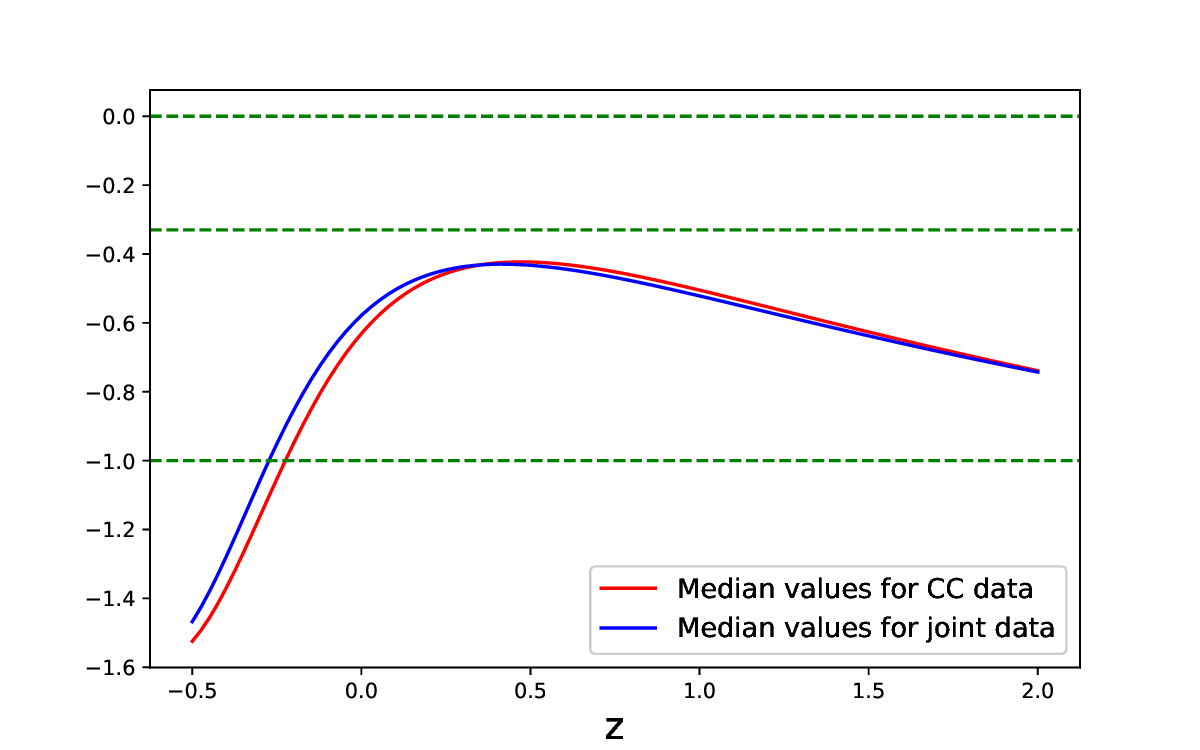}
		\caption{ {\bf{For Model-I:}} EoS parameter ($\omega_{de}$) versus $z$.}
		\label{fig:8}
	\end{figure}
\end{center}

\begin{center}
	\begin{figure}
		\includegraphics[width=16cm, height=6cm]{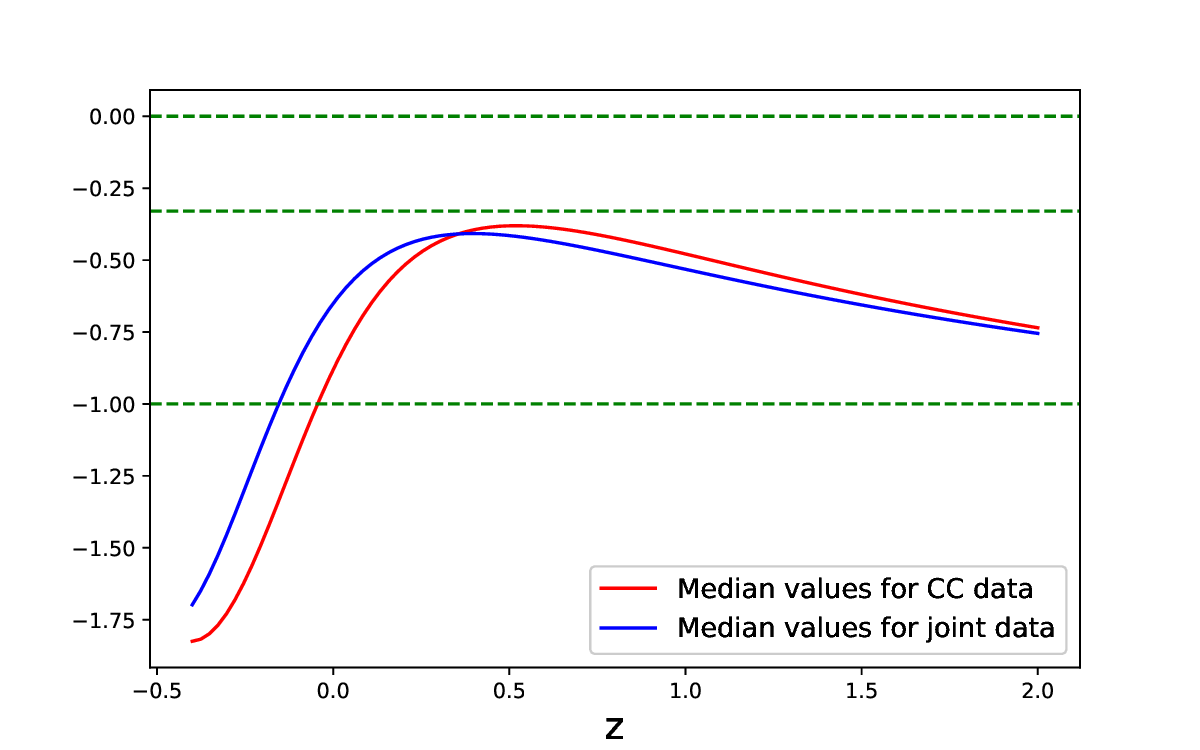}
		\caption{ {\bf{For Model-II:}} EoS parameter ($\omega_{de}$) versus $z$.}
		\label{fig:9}
	\end{figure}
\end{center}

In order to analyze the dynamical behavior of the cosmological framework, we need a functional form of $f(T)$. We adopt a well-motivated form of $f(T)$ given as~\cite{yang2011new}:
\begin{equation}{\label{34}}
	f(T) = \frac{\alpha T_0 \left( \frac{T^2}{T_0^2} \right)^\beta}{\left( \frac{T^2}{T_0^2} \right)^\beta + 1} + T,	
\end{equation}
where $\alpha$ and $\beta$ are the model parameters with $T_0 = -6H_0^2$ representing the torsion scalar at the present epoch. For brevity, we denote the first derivative by $f_T$ and the time derivative of $f_{T}$ by $\dot{f_{T}}$, which can be expressed as,
\begin{equation*}
f_T = \frac{2\alpha \beta T_0 \left( \frac{T^2}{T_0^2} \right)^\beta}{T \left( \left( \frac{T^2}{T_0^2} \right)^\beta + 1 \right)^2} + 1,
\end{equation*}
\begin{equation*}
\dot{f_{T}} = -24\alpha \beta H \dot{H} T_{0}\left(\frac{(2\beta-1)\left( \frac{T^2}{T_0^2} \right)^\beta}{T^{2} \left(\left( \frac{T^2}{T_0^2} \right)^\beta+1\right)^{2}}-\frac{4\beta \left( \frac{T^2}{T_0^2} \right)^{2\beta}}{T^{2} \left(\left( \frac{T^2}{T_0^2} \right)^\beta+1\right)^{3}}\right).
\end{equation*}
%%%%%%%%%%%%%%%%%%%%%%%%%%%%%%%%%%%%%%%%%%%%%%%%%%%%%%%%%%%%%%%%%%%%%%%%%%
\\  \\
We examine the physical behavior of fundamental quantities, namely the dark energy density and pressure. By using Eqs.~(\ref{17}), (\ref{18}) and (\ref{23}), the expressions for the dark energy density $(\rho_{de})$ and dark energy pressure $(p_{de})$ for Model-I are derived as,
%%%%%%%%%%%%%%%%%%%%%%%%%%%%%%%%%%%%%%%%%%%%%
\begin{equation}{\label{35}}
\rho_{de}(z) = \left(\frac{2\alpha \beta T_{0}\left(E_{1}\right)^{4\beta}}{\left(\left(E_{1}\right)^{4\beta}+1\right)^{2}}\right)-\left(\frac{\alpha T_{0}\left(E_{1}\right)^{4\beta}}{2\left(E_{1}\right)^{4\beta}+2}\right)-   3H_{0}^{2} \left(E_{1}\right)^{2}       \qquad \qquad \qquad \qquad\qquad    \text{(for Model-I)}
\end{equation}
%%%%%%%%%%%%%%%%%%%%%%%%%%%%%%%%%%%%%%%%%%%%%%%%%%%%%%%%
\begin{equation} {\label{36}}
\begin{split}
p_{de}(z)=&\frac{1}{2}  \left(\frac{\alpha T_{0}\left(E_{1}\right)^{4\beta}}{\left(E_{1}\right)^{4\beta}+1}  -   6H_{0}^{2} \left(E_{1}\right)^{2}\right) +2H_{0}^{2} \left(E_{1}\right)^{2} \left(3-\gamma \left[\frac{\left(\zeta(1+z)\right)^{\gamma}-1}{1+\left(\zeta(1+z)\right)^{\gamma}}\right]\right)  \left(\frac{2\alpha \beta \left(E_{1}\right)^{4\beta-2}}{\left(\left(E_{1}\right)^{4\beta}+1\right)^{2}}+1\right) \\& + \frac{4}{3} \alpha \beta \gamma T_{0} \left( \frac{\left(\zeta(1+z)\right)^{\gamma}-1}{1+\left(\zeta(1+z)\right)^{\gamma}}\right) \left( \left(\frac{(2\beta-1)\left(E_{1}\right)^{4\beta}}{\left(\left(E_{1}\right)^{4\beta}+1\right)^{2}}\right) - \left(\frac{4\beta \left(E_{1}\right)^{8\beta}}{\left(\left(E_{1}\right)^{4\beta}+1\right)^{3}}\right)   \right)    \quad  \text{(for Model-I)}
\end{split}	
\end{equation} 
where $E_{1}$ = $(1+\zeta^{\gamma})^{-2} (1+z)^{-\gamma} \left[1+(\zeta (1+z))^{\gamma} \right]^{2}$.                \\  \\
%%%%%%%%%%%%%%%%%%%%%%%%%%%%%%%%%%%%%%%%%%%%%%%%%%%%%%%%%%%%%%%
By employing Eqs.~(\ref{17}), (\ref{18}) and (\ref{24}), the expressions for the dark energy density $(\rho_{de})$ and dark energy pressure $(p_{de})$ for Model-II are obtained as,
%%%%%%%%%%%%%%%%%%%%%%%%%%%%%%%%%%%%%%%%%%%%%%%%%%%
\begin{equation}{\label{37}}
\rho_{de}(z) = \left(\frac{2\alpha \beta T_{0}\left(E_{2}\right)^{4\beta}}{\left(\left(E_{2}\right)^{4\beta}+1\right)^{2}}\right)-\left(\frac{\alpha T_{0}\left(E_{2}\right)^{4\beta}}{2\left(E_{2}\right)^{4\beta}+2}\right)-   3H_{0}^{2} \left(E_{2}\right)^{2}       \qquad \qquad \qquad \qquad\qquad    \text{(for Model-II)}
\end{equation}
%%%%%%%%%%%%%%%%%%%%%%%%%%%%%%%%%%%%%%%%%%%%%%%%%%%%
%%%%%%%%%%%%%%%%%%%%%%%%%%%%%%%%%%%%%%%%%%%%%%%%%%%%%%%%
\begin{equation} {\label{38}}
\begin{split}
p_{de}(z)=&\frac{1}{2}  \left(\frac{\alpha T_{0}\left(E_{2}\right)^{4\beta}}{\left(E_{2}\right)^{4\beta}+1}  -   6H_{0}^{2} \left(E_{2}\right)^{2}\right) +2H_{0}^{2} \left(E_{2}\right)^{2} \left(3+ \left[2\gamma-\frac{3\gamma \left(\zeta(1+z)\right)^{2\gamma}}{1+\left(\zeta(1+z)\right)^{2\gamma}}\right]\right)  \left(\frac{2\alpha \beta \left(E_{2}\right)^{4\beta-2}}{\left(\left(E_{2}\right)^{4\beta}+1\right)^{2}}+1\right) \\& - \frac{4}{3} \alpha \beta \gamma T_{0} \left( 2-\frac{3 \left(\zeta(1+z)\right)^{2\gamma}}{1+\left(\zeta(1+z)\right)^{2\gamma}}\right)  \left( \left(\frac{(2\beta-1)\left(E_{2}\right)^{4\beta}}{\left(\left(E_{2}\right)^{4\beta}+1\right)^{2}}\right) - \left(\frac{4\beta \left(E_{2}\right)^{8\beta}}{\left(\left(E_{2}\right)^{4\beta}+1\right)^{3}}\right)   \right)          \text{(for Model-II)}
\end{split}	
\end{equation} 
where $E_{2}$ = $(1+\zeta^{2\gamma})^{-\frac{3}{2}} (1+z)^{-2\gamma} \left[1+(\zeta (1+z))^{2\gamma} \right]^{\frac{3}{2}} $.  \\  \\ 
%%%%%%%%%%%%%%%%%%%%%%%%%%%%%%%%%%%%%%%%%%%%%%%%%%%%%%%%%%%%
The evolutionary patterns of energy density ($\rho_{de}$) and pressure ($p_{de}$) are illustrated in Figure (\ref{fig:6}) for Model-I and in Figure (\ref{fig:7}) for Model-II. In both models, for the set of constrained parameters, the energy density remains positive throughout the entire cosmic evolution. From high to low redshifts, the pressure stays negative, displaying a characteristic transition to enhanced negative values in the late universe as dark energy comes to dominate the expansion history. As the universe undergoes the shift from decelerated to accelerated expansion, the energy density consistently retains its positive nature, while the pressure becomes increasingly negative due to the growing dominance of dark energy. The universe may be expanding more quickly in this late era due to the pressure’s negative nature. These observations are in agreement with the universe's ongoing accelerated expansion. The values of the model parameters $\alpha$ = $240.3$ and $\beta$ = $0.37$ are chosen to ensure consistency with observational constraints and to produce physically viable cosmological behavior. In particular, these values lead to a Hubble expansion history that remains compatible with late-time cosmic acceleration and ensures that fundamental physical requirements, such as positive energy density and physically viable pressure evolution, are consistently satisfied. Furthermore, the selected parameter set avoids divergences and ensures the stability of the models throughout the considered redshift range. Therefore, these values are adopted for illustrative and graphical analysis.
%%%%%%%%%%%%%%%%%%%%%%%%%%%%%%%%%%%%%%%%%%%%%%%%%%%%%%%
\\ \\
In cosmological studies, the Equation of State (EoS) parameter plays an essential role in describing the nature of dark energy. It defines the relationship between the pressure and the energy density of the cosmic fluid. The EoS parameter is defined mathematically as $\omega$ = $\frac{p}{\rho}$. The value of the EoS parameter provides important information about different evolutionary phases of the universe. In particular, $\omega=0$ corresponds to pressureless matter (dust), $\omega=\frac{1}{3}$ describes the radiation-dominated era, and $\omega=-1$ represents vacuum energy associated with the de Sitter phase. Moreover, the accelerated expansion of the universe occurs when the EoS parameter satisfies $\omega <$ $-\frac{1}{3}$. This region includes the phantom regime ($\omega < -1 $) as well as the quintessence regime ($-1 < \omega < -\frac{1}{3}$). 
%%%%%%%%%%%%%%%%%%%%%%%%%%%%%%%%%%%%%%%%%%%%%%%%%%%%%%%%%%%
The dark energy EoS parameter $\omega_{de}$ for Model-I is obtained by utilizing equations (\ref{35}) and (\ref{36}):
\small{
\begin{equation}{\label{39}}
\omega_{de}(z)= -1 + \frac{4\gamma H_{0}^{2}  \left( \frac{\left(\zeta(1+z)\right)^{\gamma}-1}{1+\left(\zeta(1+z)\right)^{\gamma}}\right) \left(4\alpha \beta \left(\frac{(2\beta-1)\left(E_{1}\right)^{4\beta}}{\left(\left(E_{1}\right)^{4\beta}+1\right)^{2}} - \frac{4\beta \left(E_{1}\right)^{8\beta}}{\left(\left(E_{1}\right)^{4\beta}+1\right)^{3}}\right) + \left(\frac{2\alpha \beta \left(E_{1}\right)^{4\beta}}{\left(\left(E_{1}\right)^{4\beta}+1\right)^{2}}+\left(E_{1}\right)^{2}\right)     \right) }                        {\left(\frac{24 \alpha \beta H_{0}^{2}  \left(E_{1}\right)^{4\beta}}{\left(\left(E_{1}\right)^{4\beta}+1\right)^{2}}\right) + \left(\frac{\alpha T_{0}\left(E_{1}\right)^{4\beta}}{\left(E_{1}\right)^{4\beta}+1}\right)+   6H_{0}^{2} \left(E_{1}\right)^{2}}                \quad           \text{(for Model-I)}
\end{equation}}
where $E_{1}$ = $(1+\zeta^{\gamma})^{-2} (1+z)^{-\gamma} \left[1+(\zeta (1+z))^{\gamma} \right]^{2}$.                \\  \\
Further, using equations (\ref{37}) and (\ref{38}), we obtain $\omega_{de}$ for model-II as,
\small{
\begin{equation}{\label{40}}
\omega_{de}(z)= -1 - \frac{4\gamma H_{0}^{2}  \left( 2- \frac{3 \left(\zeta(1+z)\right)^{2\gamma}}{1+\left(\zeta(1+z)\right)^{2\gamma}}\right) \left(4\alpha \beta \left(\frac{(2\beta-1)\left(E_{2}\right)^{4\beta}}{\left(\left(E_{2}\right)^{4\beta}+1\right)^{2}} - \frac{4\beta \left(E_{2}\right)^{8\beta}}{\left(\left(E_{2}\right)^{4\beta}+1\right)^{3}}\right) + \left(\frac{2\alpha \beta \left(E_{2}\right)^{4\beta}}{\left(\left(E_{2}\right)^{4\beta}+1\right)^{2}}+\left(E_{2}\right)^{2}\right)     \right)}{\left(\frac{24 \alpha \beta H_{0}^{2}  \left(E_{2}\right)^{4\beta}}{\left(\left(E_{2}\right)^{4\beta}+1\right)^{2}}\right) + \left(\frac{\alpha T_{0}\left(E_{2}\right)^{4\beta}}{\left(E_{2}\right)^{4\beta}+1}\right)+   6H_{0}^{2} \left(E_{2}\right)^{2}}            \quad           \text{(for Model-II)}
\end{equation}}
where $E_{2}$ = $(1+\zeta^{2\gamma})^{-\frac{3}{2}} (1+z)^{-2\gamma} \left[1+(\zeta (1+z))^{2\gamma} \right]^{\frac{3}{2}} $.  \\  \\ 
The graphical evolution of the EoS parameter for both proposed models is illustrated in Figures (\ref{fig:8}) and (\ref{fig:9}). For Model-I, the EoS values at the current epoch ($z=0$) are determined to be $\omega_{de}=-0.6313$ for the CC dataset and $\omega_{de}=-0.5785$ for the Joint dataset. Similarly, for Model-II, the EoS parameter values are $\omega_{de}=-0.8810$ and $\omega_{de}=-0.6508$ for the CC and Joint datasets, respectively. Analysis of the $\omega_{de}$ trajectories indicates that both models exhibit quintessence-like dark energy behavior across both datasets during the current epoch. This behavior suggests a dynamic dark energy component that accounts for the observed late-time cosmic acceleration. Furthermore, for the median values of the model parameters, the universe is projected to eventually cross the cosmological constant boundary ($\omega =-1$), transitioning into a phantom-type dark energy phase at late times implies an increasing dark energy density as the universe expands.
%%%%%%%%%%%%%%%%%%%%%%%%%%%%%%%%%%%%%%%%%%%%%%%%%%%%%%%
%%%%%%%%%%%%%%%%%%%%%%%%%%%%%%%%%%%%%%%%%%%%%%%%%%%%%%%%%%%
%%%%%%%%%%%%%%%%%%%%%%%%%%%%%%%%%%%%%%%%%%%%%%%%%%%%%%%
%%%%%%%%%%%%%%%%%%%%%%%%%%%%%%%%%%%%%%%%%%%%%%%%%%%%%%%%%%%
\subsection{Energy Conditions}\label{sec:5.2}
\begin{center}
	\begin{figure}
		\includegraphics[width=18cm, height=7cm]{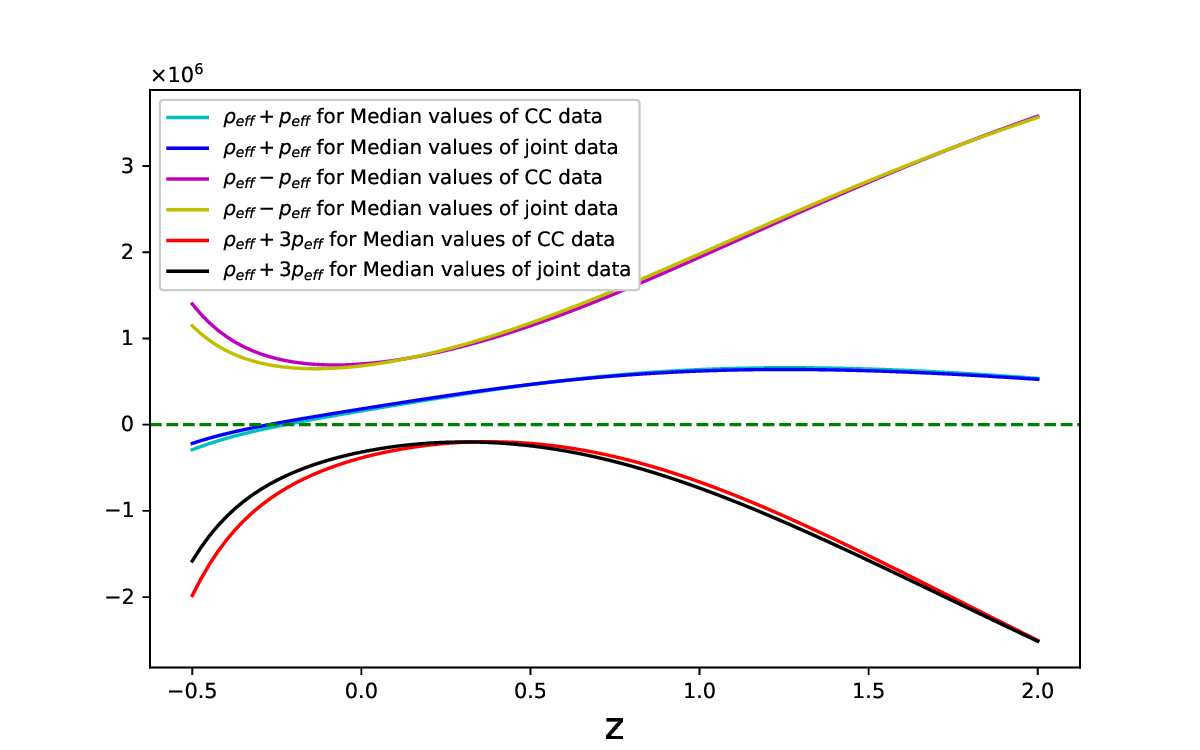}
    	\caption{{\bf{For Model-I:}} The components of energy conditions versus $\mathit{z}$.  }
		\label{fig:10}
	\end{figure}
\end{center}

\begin{center}
	\begin{figure}
		\includegraphics[width=18cm, height=7cm]{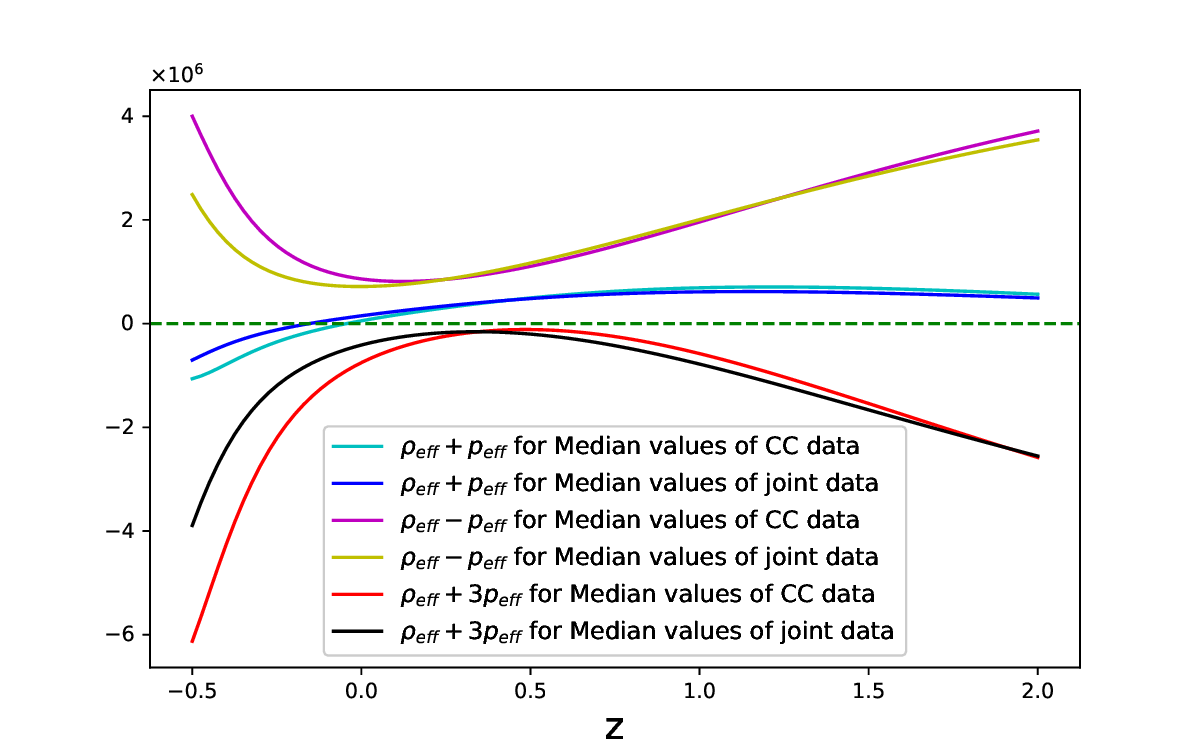}
	    \caption{{\bf{For Model-II:}} The components of energy conditions versus $\mathit{z}$.  }
		\label{fig:11}
	\end{figure}
\end{center}

To ensure the physical viability of our proposed models, we employ point-wise energy conditions based on the stress-energy tensor. Following the frameworks established in \cite{visser1997energy,lalke2024cosmic, singh2022lagrangian}, these conditions are expressed as:
%%%%%%%%%%%%%%%%%%%%%%%%%%%%%%%%%%%%%%%%%%%%%%%%%%%%%%%%%%%%
\begin{itemize}
    \item \textbf{Null Energy Condition (NEC):} Defined by the inequality $\rho_{eff} + p_{eff} \geq 0$, the NEC requires that the sum of the energy density and the effective pressure remains non-negative.
	
    \item \textbf{Weak Energy Condition (WEC):} The WEC stipulates that $\rho_{eff} \geq 0$ and  $ \rho_{eff} + p_{eff} \geq 0$, ensuring that both the energy density and its sum with the effective pressure remain non-negative.
	
    \item \textbf{Dominant Energy Condition (DEC):} This condition stipulates that the energy density must be non-negative and greater than or equal to the absolute magnitude of the pressure, expressed as $ \rho_{eff}  \geq |p_{eff}| $.
	
	\item \textbf{Strong Energy Condition (SEC):} The SEC requires that the energy density and pressure satisfy the inequalities: $\rho_{eff} + p_{eff} \geq 0$ and $\rho_{eff} + 3p_{eff} \geq 0$.
	
\end{itemize}
%%%%%%%%%%%%%%%%%%%%%%%%%%%%%%%%%%%%%%%%%%%%%%%%%%%%%%%%%%%%
The Strong Energy Condition (SEC) is characterized by the inequality $\rho_{eff} + 3p_{eff} \geq 0$, a condition fundamentally linked to the Raychaudhuri equation for determining the acceleration or deceleration of cosmic expansion~\cite{singh2023homogeneous}. Modern cosmological observations indicate that the SEC is violated globally, a phenomenon that underpins the late-time accelerated expansion of the universe. Observational constraints further indicate that the violation of the SEC has been a persistent feature of cosmic evolution, spanning from the epoch of galaxy formation to the current epoch. This sustained violation is essential to account for the observed late-time acceleration, suggesting the dominance of a dark energy component with a sufficiently negative equation of state. Consequently, an accelerating expansion rate, indicative of repulsive gravitational effects, is expected when $\rho_{eff} + 3p_{eff} < 0$. This violation points toward the existence of dark energy components with negative pressure that exhibit anti-gravitational properties. It is important to note that the SEC is formally composed of two inequalities; however, the violation of either individual constraint is sufficient to invalidate the SEC as a whole~\cite{myrzakulov2023quintessence,lalke2024cosmic, singh2023homogeneous}.
\vspace{0.2cm}\\
Figures (\ref{fig:10}) and (\ref{fig:11}) illustrate the evolution of all energy conditions for Model-I and Model-II, respectively. The graphical analysis demonstrates that both models satisfy the Null Energy Condition (NEC), Weak Energy Condition (WEC), and Dominant Energy Condition (DEC) from the past up to the current epoch. However, as the reconstructed models account for the observed late-time acceleration, the SEC (specifically the constraint $\rho_{eff} + 3p_{eff} > 0$) is clearly violated. Furthermore, the results indicate that the NEC is violated as the universe transitions into the phantom regime, which may also lead to the violation of both the WEC and DEC. The breach of the NEC specifically confirms the presence of phantom-type dark energy. Consequently, our analysis suggests that phantom dark energy remains a viable candidate and cannot be ruled out in both models. Additionally, the violation of the SEC ($\rho_{eff} + 3p_{eff} > 0$) persists and remains valid throughout the phantom-dominated phase.
%%%%%%%%%%%%%%%%%%%%%%%%%%%%%%%%%%%%%%%%%%%%%%%%%%%%%%%%%%%
%%%%%%%%%%%%%%%%%%%%%%%%%%%%%%%%%%%%%%%%%%%%%%%%%%%%%%%%%%%
%\newpage
%%%%%%%%%%%%%%%%%%%%%%%%%%%%%%%%%%%%%%%%%%%%%%%%%%%%%%%%%%%%%%%%%%%%
\subsection{Statefinder Diagnostic}\label{sec:5.3}
\begin{center}
	\begin{figure}
		\includegraphics[width=16cm, height=7cm]{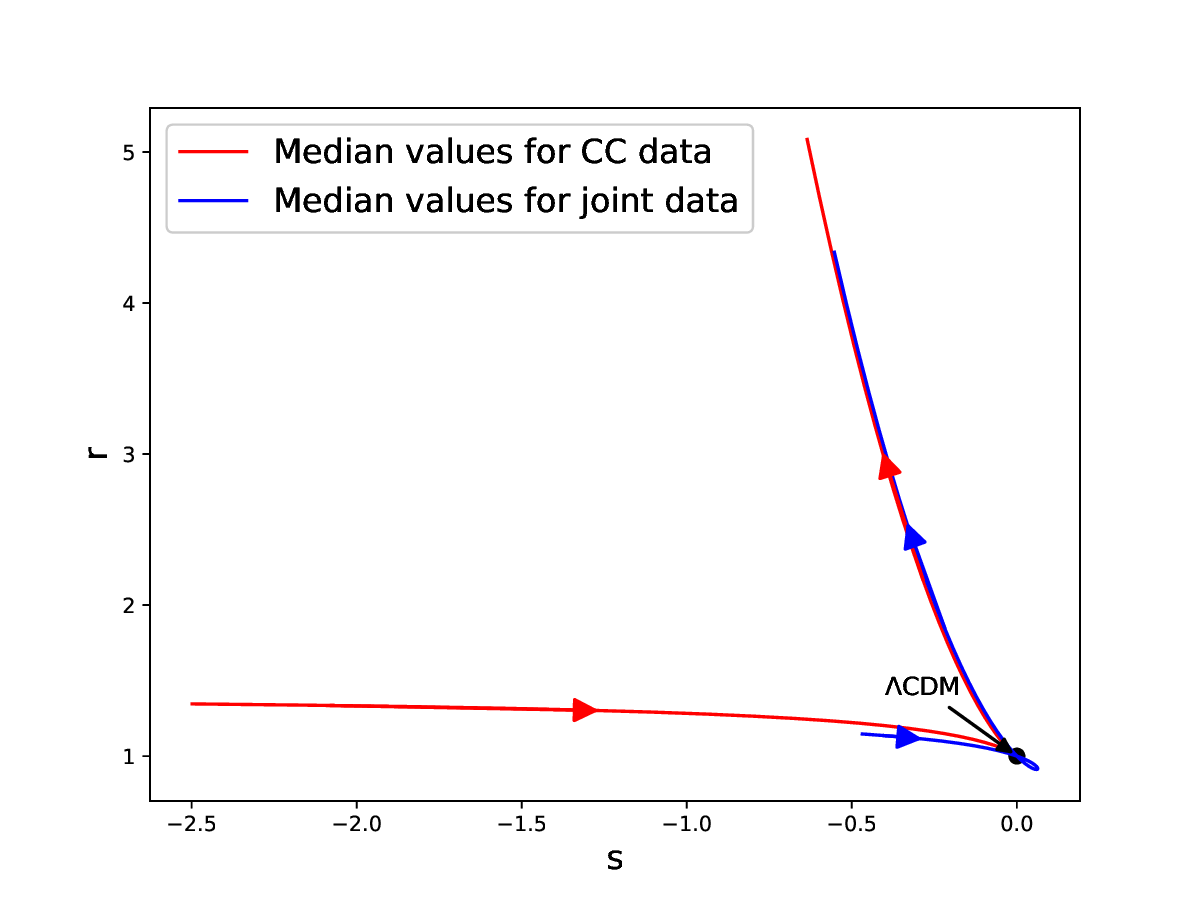}
		\caption{{\bf{For Model-I:}} Variation of $s$ and $r$ plane.  }
		\label{fig:12}
	\end{figure}
\end{center}

\begin{center}
	\begin{figure}
		\includegraphics[width=16cm, height=7cm]{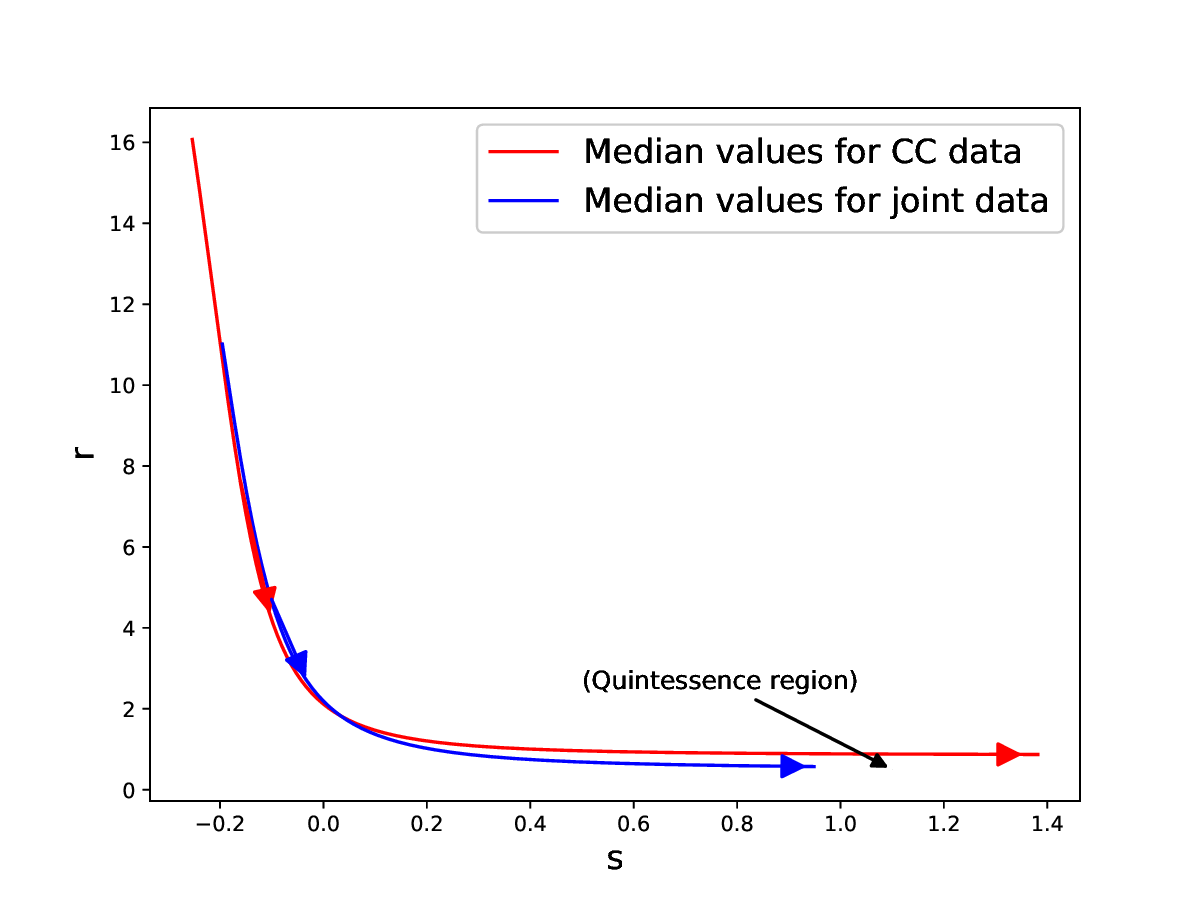}
		\caption{{\bf{For Model-II:}} Variation of $s$ and $r$ plane.  }
		\label{fig:13}
	\end{figure}
\end{center}

The role of geometric parameters in elucidating the cosmological dynamics of a model is well recognized. However, to effectively probe alternative dark energy candidates that deviate from the standard $\Lambda$CDM framework, it becomes essential to look beyond the basic descriptions provided by the Hubble parameter ($H$) and deceleration parameters ($q$). Consequently, higher-order derivatives of the scale factor $a(t)$ are employed as refined mathematical tools to detect the subtle dynamical differences between various dark energy models. To this end, the statefinder diagnostic-represented by the geometric pair $\left\{r, s\right\}$~\cite{Sahni2003}, serves as a powerful discriminatory tool, allowing for the precise classification and differentiation of various dark energy models based on their evolutionary paths. The statefinder parameters $\left\{r, s\right\}$ are defined as:
\begin{equation}{\label{41}}
r=\frac{\dddot a}{aH^{3}}, \quad s= \frac{r-1}{3(q-\frac{1}{2})}, \quad \text{where} \quad q\neq \frac{1}{2}.
\end{equation}
A wide array of competing dark energy models discussed in the literature can be effectively characterized within the $\left\{r, s\right\}$ diagnostic plane. Within this framework, the Chaplygin gas model occupies the region defined by $r >1$ and $s <0$. The $\Lambda$CDM paradigm is uniquely represented by the fixed point ($r=1$, $s=0$), whereas quintessence models are situated in the $r <1$ and $s >0$ domain. Furthermore, the holographic dark energy model is identified by the fixed point ($r=1$, $s = \frac{2}{3}$) and the standard Cold Dark Matter (SCDM) scenario is located at ($r=1$, $s=1$) in the r-s plane.
\\ \\ 
The evolutionary trajectories in the $\left\{r, s\right\}$ diagnostic plane for Model-I and Model-II are displayed in Figures (\ref{fig:12}) and (\ref{fig:13}), respectively. For Model-I, Figure (\ref{fig:12}) depicts a trajectory that initiates within the Chaplygin gas regime at early cosmic times, subsequently intersects the $\Lambda$CDM fixed point and eventually converges toward a unified framework that integrates both dark matter and dark energy components. In contrast, Figure (\ref{fig:13}) reveals that the statefinder parameters for Model-II emerge from the Chaplygin gas domain and after passing in close proximity to the $\Lambda$CDM benchmark, progress toward the quintessence region at the present epoch for both observational datasets. Such evolutionary features of the statefinder trajectories have also been reported in the literature~\cite{fei2013statefinder}.
%%%%%%%%%%%%%%%%%%%%%%%%%%%%%%%%%%%%%%%%%%%%%%%%%%%%%%%%%%%
%%%%%%%%%%%%%%%%%%%%%%%%%%%%%%%%%%%%%%%%%%%%%%%%%%%%%%%%%%%
%%%%%%%%%%%%%%%%%%%%%%%%%%%%%%%%%%%%%%%%%%%%%%%%%%%%%%%%%%%
%%%%%%%%%%%%%%%%%%%%%%%%%%%%%%%%%%%%%%%%%%%%%%%%%%%%%%%%%%%
\subsection{Age of the Universe}\label{sec:5.4}
The evolution of the cosmic age $t(z)$ with  redshift $ \mathit{z} $ for the cosmological model is presented in \cite{tong2009cosmic}.
\begin{equation} {\label{42}}
t(z) = \int_{z}^{\infty} \frac{dz}{(1+z)H(z)}.
\end{equation}
The current age of the universe, $t_{0}$, is determined by numerically evaluating the integral of the Hubble parameters $H(z)$ (from Eqs. \ref{23} and \ref{24}) at the present epoch ($ z = 0 $). For Model-I, we obtain $t_{0} =13.27$ Gyr (CC dataset) and $t_{0} =13.38$ Gyr (joint dataset). For Model-II, the present age is found to be $t_{0} =13.63$ Gyr and $t_{0} =14.01$ Gyr for the CC and joint datasets, respectively. Notably, these findings are close to the present age values of the $\Lambda$CDM model as reported in Planck results~\cite{2020A&A...641A...6P} and align with other observational study~\cite{Valcin2020}.
%%%%%%%%%%%%%%%%%%%%%%%%%%%%%%%%%%%%%%%%%%%%%%%%%%%%%%%%%%%%%%%%%
\section{Conclusions}\label{sec:6}
The cosmic evolution of a spatially flat FLRW universe is investigated within the $f(T)$ gravity framework, utilizing a well-motivated functional form of $f(T)$. In this work, we analyze the cosmological dynamics by considering a specific parameterized form of the Hubble parameter $H(t)$, which leads to two models, referred to as Model-I and Model-II (as summarized in Table (\ref{table:1})). The median values of the model parameters were determined using the Bayesian statistical approach through the MCMC analysis implemented in the emcee package. The parameters were constrained using the cosmic chronometer (CC) dataset as well as the joint (CC + Pantheon) dataset. The obtained results are summarized in Table(\ref{table:2}) and Table(\ref{table:3}). The best-fit curve of the Hubble parameter $H(z)$ demonstrate the compatibility of the both models with the measured cosmic chronometer observations.
\vspace{.2cm}\\
We further analyzed the evolutionary dynamics of the universe through various cosmological parameters. In particular, the behavior of the deceleration parameter has been investigated in detail. The evolution curves of the deceleration parameter (Figures (\ref{fig:1}) and (\ref{fig:2})) clearly illustrate the transition of the universe from an earlier decelerated phase to the present accelerated phase of cosmic expansion. For Model-I, the present value of the deceleration parameter is obtained as $q_{0} = -0.6151$ (CC) and $q_{0} = -0.5785$ (joint dataset), while the transition redshift is found to be $z_{t} = 0.6261$ (CC) and $z_{t} = 0.6263$ (joint dataset). Similarly, for Model-II the present values of the deceleration parameter are $q_{0} = -0.8758$ (CC) and $q_{0} = -0.6508$ (joint dataset), with the transition redshift $z_{t} = 0.6036$ (CC) and $z_{t} = 0.6039$ (joint dataset). The negative values of the present deceleration parameter $q_{0}$ confirms that the universe is currently ($z = 0$) undergoing accelerated expansion. These findings are consistent with the observed late-time acceleration and the expanding behavior of the universe.
\vspace{.2cm}\\
Additionally, the behavior of the physical parameters associated with dark energy has been examined. The energy density of dark energy $(\rho_{de})$ remains positive throughout the cosmic evolution, whereas the pressure $(p_{de})$ remains negative and becomes increasingly dominant at late times due to the dominance of dark energy. This negative pressure component is responsible for driving the accelerated expansion of the universe in the present epoch.
Furthermore, the evolution of the EoS parameter ($\omega_{de}$) for dark energy has been analyzed and depicted in the Figures (\ref{fig:8}) and (\ref{fig:9}). For Model-I, the present values of the EoS parameter are $\omega_{DE} = -0.6313$ (CC) and $\omega_{DE} = -0.5785$ (joint dataset). Similarly, for Model-II the present values are $\omega_{DE} = -0.8810$ (CC) and $\omega_{DE} = -0.6508$ (joint dataset). These values indicate that both models lie within the quintessence region at the present epoch. However, the future evolution of the EoS parameter reveals a crossing of the cosmological constant boundary $\omega=-1$, indicating a transition to a phantom regime and implying a quintom-like behavior of dark energy.
\vspace{.2cm}\\
Finally, we estimated the present age of the universe for both models. For Model-I, the age of the universe is found to be $t_{0} = 13.27$ Gyr and $t_{0} = 13.38$ Gyr for the CC and joint datasets, respectively. For Model-II, the corresponding values are $t_{0} = 13.63$ Gyr (CC) and $t_{0} = 14.01$ Gyr (joint dataset). These values fall within the acceptable observational range for the cosmic age. In conclusion, the overall analysis indicates that both proposed models successfully describe the late-time accelerated expansion of the universe and remain consistent with the available observational datasets. Therefore, the considered parametric framework within $f(T)$ gravity provides a viable and physically consistent description of the present cosmic dynamics.  
%%%%%%%%%%%%%%%%%%%%%%%%%%%%%%%%%%%%%%%%%%%%%
\section*{\textbf{Acknowledgements}}
One of the authers, G. P. Singh gratefully acknowledges the support provided by the Inter-University Centre for Astronomy and Astrophysics (IUCAA), Pune, India, under the Visiting Associateship Programme.
%%%%%%%%%%%%%%%%%%%%%%%%%%%%%%%%%%%%%%%%%%%%%%%%%%%
%\section*{\textbf{Data Availability statement}}
%All the publicly available data sources has been cited in the manuscript.
%%%%%%%%%%%%%%%%%%%%%%%%%%%%%%%%%%%%%%%%%%%%%%%%%%
%\section*{\textbf{Declaration of competing interest}}
%The authors declare that they have no known competing financial interests or personal relationships that could have appeared to influence the work reported in this paper. 

\end{document}